\documentclass[twocolumn]{article}
\usepackage{stfloats}
\usepackage{color}
\usepackage{subfigure}
\usepackage{graphicx}
\usepackage{amsmath,amssymb,amsfonts}
\usepackage{amsthm}
\usepackage{bm}
\usepackage[hyphens]{url}
\usepackage{listings}
\usepackage{multirow}
\usepackage{authblk}

\usepackage[linesnumbered,ruled,vlined]{algorithm2e}

\theoremstyle{plain}
\newtheorem{thm}{Theorem}
\theoremstyle{definition}
\newtheorem{defn}[thm]{Definition}

\usepackage[noadjust]{cite}

\newif\ifdraft
\draftfalse
\ifdraft
 \newcommand{\eunsungnote}[1]{ {\textcolor{red} { ***Eunsung: #1 }}}
 \newcommand{\rajnote}[1]{ {\textcolor{blue} { ***Raj: #1 }}}
 \newcommand{\samnote}[1]{ {\textcolor{cyan} { ***Sam: #1 }}}
 \newcommand{\samtextold}[1]{ {\textcolor{green} { ***Sam Old Text: #1 }}}
  \newcommand{\samtextnew}[1]{ {\textcolor{magenta} { ***Sam New Text: #1 }}}

 \newcommand{\ian}[1]{ {\textcolor{green} { ***Ian: #1 }}}
\else
 \newcommand{\eunsungnote}[1]{}
 \newcommand{\rajnote}[1]{}
 \newcommand{\samnote}[1]{}
 \newcommand{\samtextold}[1]{}
 \newcommand{\samtextnew}[1]{}
 \newcommand{\ian}[1]{}
\fi
\newcommand{\ignore}[1]{}

\title{Towards Accommodating Real-time Jobs on HPC Platforms} 

\author[1]{Sam~Nickolay}
\author[2]{Eun-Sung~Jung}
\author[3]{Rajkumar~Kettimuthu}
\author[1]{Ian~Foster}
\affil[1]{The University of Chicago.(e-mail: samnickolay@uchicago.edu, foster@cs.uchicago.edu)}
\affil[2]{Hongkik University.(e-mail: ejung@hongik.ac.kr)}
\affil[3]{Argonne National Laboratory.(e-mail: kettimut@anl.gov)}

\date{}

\begin{document}
\maketitle

\begin{abstract}
Increasing data volumes in scientific experiments necessitate the use of high performance computing (HPC) resources for data analysis. In many scientific fields, the data generated from scientific instruments and supercomputer simulations must be analyzed rapidly. In fact, the requirement for quasi-instant feedback is growing. Scientists want to use results from one experiment to guide the selection of the next or even to improve the course of a single experiment. Current HPC systems are typically batch-scheduled under policies in which an arriving job is run immediately only if enough resources are available; otherwise it is queued. It is hard for these systems to support real-time jobs. Real-time jobs, in order to meet their requirements, should sometimes have to preempt batch jobs and/or be scheduled ahead of batch jobs that were submitted earlier. 
Accommodating real-time jobs may negatively impact system utilization also, especially when preemption/restart of batch jobs is involved.
We first explore several existing scheduling strategies to make real-time jobs more likely to be scheduled in due time. We then rigorously formulate the problem as a mixed-integer linear programming for offline scheduling and develop novel scheduling heuristics for online scheduling. We perform simulation studies using trace logs of Mira, the IBM BG/Q system at Argonne National Laboratory, to quantify the impact of real-time jobs on batch job performance for various percentages of real-time jobs in the workload. We present new insights gained from grouping jobs into different categories based on runtime and the number of nodes used and studying the performance of each category. Our results show that with 10\% real-time job percentages, just-in-time checkpointing combined with our heuristic can improve the slowdowns of real-time jobs by 35\% while limiting the increase of the slowdowns of batch jobs to 10\%.
\end{abstract}

\section{Introduction}
\label{sec_introduction}

Scientific instruments such as accelerators, telescopes, light sources, and colliders generate large amounts of data. Because of advances in technology, the rate and size of these data are rapidly increasing. Advanced instruments such as state-of-the-art detectors at light source facilities generate tens of terabytes of data per day, and future camera-storage bus technologies are expected to increase data rates by an order of magnitude or more. The ability to quickly perform computations on these data sets will improve the quality of science. A central theme in experimental and observational science workflows is the need for quasi-instant feedback so that the result of one experiment can guide the selection of the next. Online analysis so far has typically been done by dedicated compute resources available locally at the experimental facilities. With the massive increase in data volumes, however, the computational power required to do the fast analysis of these complex data sets often exceeds the resources available locally. Hence, many instruments are operated in a blind fashion without a quick analysis of the data to give insight into how the experiment is progressing. 

Large-scale high-performance computing (HPC) platforms and supercomputers are required in order to do on-demand processing of experimental data. Such processing will help detect problems in the experimental setup and operational methods early on and will allow for adjusting experimental parameters on the fly~\cite{thomas_towards_2015}. Even slight improvements can have far-reaching benefits for many experiments. However, building a large HPC system or having a supercomputer dedicated for this purpose is not economical, because the computation in these facilities typically is relatively small compared with the lengthy process of setting up and operating the experiments. 

We define real-time computing as the ability to perform on-demand execution. The real-time computation may represent either analysis or simulation. Recently NERSC set up a ``real-time'' queue on its new Cori supercomputer to address real-time analysis needs. It uses a small number of dedicated compute nodes to serve the jobs in the real-time queue, and it allows jobs in the real-time queue to take priority on other resources. It is also possible to preempt ``killable" jobs on these other resources.
NERSC is, however, an exception. The operating policy of most supercomputers and scientific HPC systems is not suitable for real-time computations. The systems instead adopt a batch-scheduling model where a job may stay in the queue for an indeterminate period of time. Thus, existing schedulers have to be extended to support real-time jobs in addition to the batch jobs. The main challenge of using supercomputers to do real-time computation is that these systems do not support preemptive scheduling. A better understanding of preemptive scheduling mechanisms is required in order to develop appropriate policies that support real-time jobs while maintaining the efficient use of resources. 

In this paper, we present our work on evaluating various scheduling schemes to support mixes of real-time jobs and traditional batch jobs. We perform simulation studies using trace logs of Mira, the IBM BG/Q system at Argonne National Laboratory, to quantify the impact of real-time jobs on batch job performance and system utilization for various percentages of real-time jobs in the workload. 

Parallel job scheduling has been widely studied \cite{Lifka:1995,anastas:1997,jones99scheduling,slack99,Mu'alem:2001:UPW},
and surveys \cite{Feitelson:1995:PJS,Feitelson97jobscheduling,Feitelson:2004:PJS} and evaluations \cite{performance1990,Chiang:2001:PJS,relaxed:2002,multipleq:2002} have been published.
We first explore several existing scheduling strategies to make real-time jobs more likely to be scheduled in due time. 
We then rigorously formulate the problem as a mixed-integer linear programming for offline scheduling and develop novel scheduling heuristics for online scheduling. Using Mira trace logs, we quantify the impact of real-time jobs on batch jobs performance for various percentages of real-time jobs in the workload. We present new insights gained from studying the performance of different categories of jobs grouped based on runtime and the number of nodes used. Our results show that our proposed heuristic could achieve reasonable slowdowns for the accommodated real-time jobs while limiting the increase of the slowdowns for batch jobs.

In summary, the main contributions of this work are: (1) Definition of real-time jobs on high-performance computing platforms, (2) Optimal formulation as a mixed-integer linear programming for offline scheduling of real-time jobs, (3) Adaptation and analysis of existing scheduling techniques for real-time jobs, (4) Novel sophisticated heuristics for scheduling of real-time jobs, and (5) Performance evaluation through extensive simulations using actual job logs of the Mira supercomputer.

The rest of the paper is organized as follows. Section \ref{sec_related_work} describes the background of supercomputers and parallel batch job scheduling, and the related work for real-time job scheduling on modern high-performance computing (HPC) systems. 
Section~\ref{sec_problem_statment} formally states the real-time job scheduling problem.
Section~\ref{sec_basic_scheduling} presents basic scheduling techniques adapted for the real-time job scheduling problem.
Section~\ref{sec_optimization_algorithm} presents the optimization formulation and its time complexity analysis, and Section~\ref{sec_heuristic} presents the novel heuristic algorithms.
Section~\ref{sec_evaluation} describes the experimental setup and the simulation results of our novel heuristics and comparison with the optimization formulation. We conclude this paper with a summary in Section~\ref{sec_conclusion}.

\section{Background and Related Work}\label{sec_related_work}
In this section, we describe the background of this study and do literature review related to the problem of real-time scheduling of parallel scientific jobs on production supercomputers.

\subsection{Parallel job scheduling}
Parallel jobs are usually batch jobs, which are served on parallel machines such as supercomputers
in a first-come-first-served (FCFS) fashion traditionally. Most of the current systems support backfilling to address the low utilization issue with the strict FCFS policy. 
Batch jobs are queued in one or more queues, which are maintained by a parallel job scheduler.
Parallel job scheduling has been widely studied~\cite{Feitelson:1995:PJS,Feitelson97jobscheduling,Feitelson:2004:PJS}. 
\eunsungnote{The following is related to Reviewer\#4-[2] comment.}
It includes strategies such as backfilling~\cite{Mu'alem:2001:UPW,Shmueli:2005:BLO,Lifka:1995,Srinivasan:2002:SRS},
preemption~\cite{Niu:2013:ECI,Kettimuthu:2005:SPS}, moldability~\cite{Sabin:2006:MPJ,Cirne:2000:ASP,Srinivasan:2002:ESP}
malleability~\cite{CactusWorm2001}, techniques to use distributed
resources~\cite{Ranganathan:2002:DCD,Subramani:2002:DJS}, mechanisms to handle fairness~\cite{Sabin:2005:UMS,Tang:2013:TBS}, and methods to handle inaccuracies in user runtime estimates~\cite{Tang:2013:JSA}. 
Sophisticated scheduling algorithms have been developed that can optimize resource allocation while also addressing other goals such as minimizing average slowdown~\cite{Feitelson:1997:TPP} and turnaround time.

In contrast to the conventional HPC applications, recently emerging HPC applications such as light source experiments~\cite{fpwr:2017} require on-demand processing capability to expedite the overall problem-solving time. However, few studies have been done regarding supporting real-time jobs as well as batch jobs on parallel machines. Existing HPC platforms make use of tweaks such as separate queues and dedicated resources for real-time jobs. 

A survey on scalable system scheduling~\cite{reuther_scalable_2018} compared traditional supercomputing jobs to high-performance data analytics jobs which are featured by very short-run and time-critical jobs. Accordingly, a great deal of research conducted in each job categories is presented. If we name a few outstanding works among many schedulers, the traditional HPC schedulers include PBS~\cite{henderson_job_1995}, HTCondor~\cite{litzkow_condor_1988}, and LSF~\cite{zhou_utopia_1993}, and the big data schedulers include Google Borg~\cite{verma_large-scale_2015}, Apache Hadoop YARN~\cite{vavilapalli_apache_2013}, and Apache Mesos~\cite{hindman_mesos_2011}. Even though our work has similar goals to those of the big data schedulers, our work focuses on how to run time-critical jobs on the existing batch-based HPC platforms while trying to meet the timing requirements of such time-critical jobs.
In advanced research on cloud computing, we can classify previous related studies about coping with time-critical jobs in terms of HPC/cloud platforms. There exist roughly three cases. The first case is that time-critical jobs on batch-based HPC platforms are dispatched to the cloud, which is called cloud bursting~\cite{clemente-castello_performance_2018,fox_conceptualizing_2017}. The second case is that the cloud provisions on-demand resources for all jobs,  both batch and time-critical jobs~\cite{marshall_improving_2011,delamare_spequlos:_2012,yan_tr-spark_2016}. Especially Tr-Spark~\cite{yan_tr-spark_2016} studied  checkpointing methods on Spark are studied to evict big data analytics jobs running as secondary background jobs. Lastly, the case of the most recent work closely related to ours is that the central cloud controller dynamically negotiates capacity between on-demand and batch clusters so that time-critical jobs can find required resources on the on-demand cluster without being invasive on batch jobs on the batch cluster~\cite{liu_dynamically_2018}. Our work differs from the three cases mentioned above in that we try to make the best use of the existing HPC platforms for batch jobs using sophisticated scheduling algorithms combined with checkpointing and preemption techniques.  To the best of our knowledge, there has been no work on analyzing how time-critical jobs' schedules perform when checkpointing/preemption techniques apply on existing batch-based HPC platforms and appropriate algorithms for such environments are deployed. 

In this paper, we tackle this real-time job scheduling problem from a standpoint of general mechanisms based on mathematical formulations and heuristics.
In fact, the parallel job scheduling problem whose objective is minimizing makespan is NP-hard~\cite{graham_optimization_1979}. Thus, the production parallel job schedulers and the research studies on parallel job schedulers use heuristics.
Following the notation in \cite{graham_optimization_1979}, our problem is $P|pmtn,r_j|\sum (C_j-r_j)$, which is more complicated problem than NP-hard $P|pmtn|\sum C_j$, which means a scheduling problem with $P$ identical machines where job preemption is allowed and the objective is to minimize the sum of job completion times.
Whereas few studies try to compare the heuristics with optimal algorithms, we compare the results of our novel heuristics with optimal solutions of MILP-based formulation.

\subsection{Scheduling Real-time and Batch jobs}
Although preemptive scheduling is universally used at the operating-system level to multiplex processes on single-processor systems and shared-memory multiprocessors, it is rarely used in parallel job scheduling. 
Studies of preemptive scheduling schemes have focused on their overheads and their effectiveness in reducing average job turnaround time~\cite{Motwani:1993:NS,Deng:1996:PSP,Snell:2002:PBB,Chiang:2001:PJS,Kettimuthu:2005:SPS,Leung:2010:PJS}.

Others have studied preemptive scheduling for
\emph{malleable parallel jobs}~\cite{Deng:1996:PSP,parsons97,application,processor}, in which the number of processors used to execute a job is permitted to vary dynamically over time. In practice, parallel jobs submitted to supercomputer centers are generally rigid; that is, the number of processors used to execute a job is fixed.
The work most similar to ours is SPRUCE (Special Priority and Urgent Computing Environment) \cite{trebon_2011}, which investigated mechanisms for supporting \emph{urgent} jobs such as hurricane analysis on HPC resources.
The authors define urgent computing jobs as having time-critical needs, 
such that late results are useless. 
SPRUCE considered only a basic preemptive scheduling scheme with no checkpointing and assumed that urgent jobs are infrequent. 
Our work differs in terms of both its job model and the scheduling schemes considered. Our job model assumes that jobs with real-time constraints arrive more frequently and that jobs are not a total failure even if the job timing requirements are missed. We evaluate more sophisticated preemptive scheduling schemes.

\subsection{HPC Environments and Simulation}
We evaluate the scheduling algorithms using the job logs of Mira~\cite{url_mira}, which is a IBM Blue Gene/Q supercomputer at Argonne National Laboratory. 
But without loss of generality, we assume that the proposed approaches apply to other supercomputers.
\ignore{
Mira is hierarchically structured such that a midplane is composed 512 nodes,
two midplanes belong to a rack, 16 racks make up one row and Mira is finally composed of 3 rows.
Jobs on Mira are scheduled to run on partitions that have integer multiples of 512 nodes.
The batch scheduler of Mira is the Cobalt~\cite{url_Cobalt}, which is a home-grown software at ANL and is an open-source and component-based tool. The baseline results shown in this paper is based on the Cobalt job scheduler.
}

Mira is a Blue Gene/Q system operated by the Argonne Leadership Computing Facility (ALCF) at Argonne National Laboratory \cite{url_mira}.
It was ranked ninth in the 2016 Top500 list, with peak performance at 10,066 TFlop/s. 
Mira is a 48-rack system, with 786,432 cores. 
It has a hierarchical structure connected via a 5D torus network. Nodes are grouped into midplanes, each containing 512 nodes; each rack has two midplanes. 
Partitions on Mira are composed of such midplanes. Thus, jobs on Mira are scheduled to run on partitions that have integer multiples of 512 nodes. The smallest production job on Mira occupies 512 nodes, and the largest occupies 49,152 nodes. 
The Cobalt \cite{url_Cobalt} batch scheduler used on Mira is an open-source, component-based resource management tool developed at Argonne. It was used as the job scheduler on Intrepid (the supercomputer at ALCF before Mira) and is being used in other Blue Gene systems such as Frost at the National Center for Atmospheric Research \cite{url_cobalt_FROST}.

We use the Qsim discrete event simulator \cite{url_qsim} as it can simulate all the features supported by the Cobalt scheduler.
Job scheduling behavior is triggered by job-submit(Q)/job-start(S)/job-end(E) events. 
\eunsungnote{The following is related to Reviewer\#2-[2] comment.}
The latest version of Qsim supports three versions of backfilling-based job scheduling policies: first-fit (FF) backfilling, best-fit (BF) backfilling, and shortest-job-first (SJF) backfilling \cite{Mu'alem:2001:UPW}. 
By design, Qsim supports the simulation of only batch job scheduling. In this study, we extend the Qsim simulator to support real-time job scheduling using a high-priority queue and preemption. 

\ignore{
\subsection{Price model}

\begin{itemize}
\item Literature review on price model in various contexts.
\item Price model suitable for dedicated network resources.
\end{itemize}
}

\section{Problem Statement}\label{sec_problem_statment}
We define real-time and batch jobs, describe metrics used to evaluate scheduling algorithms, and formulate the problem. 

\subsection{Job Definition}
Typically HPC platforms support only batch jobs.
To support real-time computing on HPC platforms,
we categorize jobs into either \emph{batch jobs} or \emph{real-time jobs}.
Here, \emph{real-time jobs} (RTJs) are the jobs that require immediate service while \emph{batch jobs} (BJs) require best-effort service.
RTJs are also known as low-latency jobs.
Each job will have the following attributes: submission time, start time, end time, walltime, \#nodes, and priority(batch or real-time) where submission time is when the job is submitted to a queue, start time is when the job actually started on computing resources, end time is when the job ends, wall time is a user-estimated run time, \#nodes is the number of nodes requested by the job. 
The submission time, walltime, \#nodes, and priority are given by a user submitting a job while the start time is determined by the scheduler and the end time will be set when the job is completed.

\subsection{Performance Metrics}
Commonly used metrics to evaluate parallel job scheduling algorithms are response time/turnaround time of jobs and job slowdown. The response time of a job is the elapsed time after submission until the job ends. The slowdown of a job is a relative measurement of the response time with regard to the run time of the job.
We can formally define the slowdown of a job $j$ ($SD_j$) as: 
\begin{defn}\label{def_sd}
$SD_j =\frac{(endTime_j-submissionTime_j)}{runTime_j}$
\end{defn}
In order to limit the influence of very short jobs, we define the bounded slowdown of a job $j$ ($BSD_j$) as 
\begin{defn}\label{def_bsd}
$BSD_j$\\
$~~~=\frac{waitTime_j + \max (runTime_j, Bound)}{\max (runTime_j, 10)}$\\
$~~~=\frac{(endTime_j-submittionTime_j-runTime_j) + \max (runTime_j, 10)}{\max (runTime_j, Bound)}$
\end{defn}

We set the $Bound$ to be 10 minutes.

\subsection{The Real-Time Job Scheduling Problem}\label{subsec_sched_problem}

We assume HPC platforms and the working environment with the following features.
\begin{itemize}
\item Jobs submitted to the platform are managed by a central scheduler.
\item Jobs upon arrival are run immediately only if there are enough unused resources. Otherwise, the jobs are put in the waiting queue to be scheduled to execute when resources become available.
\item All jobs are rigid, which means the number of nodes allocated to a certain job does not change over time.
\end{itemize}
Our goal is to keep the slowdown of RTJs as close to 1 as possible (in other words, keep the wait time of RTJs as close to 0 as possible). At the same time, we would like to minimize the slowdown (and the wait time) of BJs. 


\ignore{
\subsection{Additional Consideration on Incentive Pricing Policy}
It is obvious that a RTJ consumes more resources than a BJ when two jobs are all same except the job type since a RTJ may preempt running BJs or hold available resources while waiting for running BJs to finish.
For such reasons, it is fair that a RTJ is charged at a higher price than a BJ, which are delayed due to RTJ priority scheduling.
In Section~\ref{sec_optimization_algorithm}, we formally analyze the fair way to charge RTJs and give incentives to BJs.
}

\section{Basic Scheduling Techniques} \label{sec_basic_scheduling}
We present and analyze five basic scheduling schemes adapted to accommodate RTJs in addition to the traditional BJs. 
Ideally, malleability, where the number of nodes of a job can dynamically shrink or expand, can be used to accommodate RTJs by adjusting nodes of running batch jobs. However, the production system such as Mira does not support such functions, and we, therefore, decided not to include malleability as basic techniques.
This section is the summary of our previous work\cite{wang_supporting_2017} to help understand the baseline techniques.

\subsection{High-Priority Queue-Based Scheduling}
\eunsungnote{This is related to Reviewer1-6 comment}
The simplest solution for RTJ scheduling is to use a separate queue for RTJs.
RTJs are enqueued in a \emph{high-priority queue}, whereas 
BJs are enqueued in a \emph{normal queue}. 
The scheduler gives priority to the jobs in the high-priority queue and blocks all the jobs in the normal queue until all the jobs in the high-priority queue are scheduled. 

\subsection{Preemptive 
Scheduling with Checkpointing}
In the preemptive scheduling schemes, if not enough resources are available to schedule a RTJ, the scheduler selects a partition for the RTJ that maximizes system utilization,  preempts any BJs running on this partition, and schedules the RTJ.  
It then resubmits those preempted BJs to the normal queue for later restart/resume. The overhead introduced by preemption impacts the jobs that are preempted as well as the system utilization.

\begin{table}[ht] \renewcommand{\arraystretch}{1.1}
	\caption{Summary of Preemptive Scheduling Schemes} %
	\centering
	\begin{tabular}{ cl }
		\hline 
		Scheme & Description   \\[0.5ex]
		\hline \hline
		NO-CKPT & \emph{No} checkpointing\\
		SYS-CKPT & System level checkpointing\\
		APP-CKPT & Application/Library level checkpointing\\
        JIT-CKPT & \emph{Just-in-time} checkpointing\\
        \hline
	\end{tabular}
	\label{table:ckpt_scheme}
\end{table} 

Checkpointing can help reduce the 
overhead of preemption, but checkpointing does not come for free. Checkpointing's impact on job runtime and system utilization needs to be accounted for as well. 
There can be four preemptive scheduling schemes as in Table~\ref{table:ckpt_scheme}.
NO-CKPT denotes no checkpointing, SYS-CKPT denotes periodic checkpointing by OS kernel/hardware\rajnote{what is this system? scheduler}\eunsungnote{from [8] in JSSPP}, 
APP-CKPT denotes checkpointing by the application (job) itself, and JIT-CKPT is checkpointing by the scheduler\rajnote{same as the comment above}\eunsungnote{replace by scheduler} right before the job is preempted.
We define the following terms to capture the overhead introduced by the preemptive scheduling schemes:
$t^{j}_{ckpt}, t^{j}_{pre}, chr^{j}_{ckpt}, chr^{sys}_{ckpt},$ and  $chr^{sys}_{pre}$. Here $t^{j}_{ckpt}$ and $t^{j}_{pre}$ are the additional time incurred for $job\ j$ due to checkpointing overhead and preemption overhead, respectively; $chr^{sys}_{ckpt}$ and $chr^{sys}_{pre}$ are the core-hours lost by the system due to checkpointing overhead and preemption overhead, respectively; and $chr^{j}_{ckpt}$  is core-hours lost by $job\ j$ due to checkpointing overhead.

{\bf NO-CKPT:} 
In NO-CKPT, 
no system- or application-level checkpointing occurs. Thus, the preempted jobs have to be restarted from the beginning. Equations~\ref{eq_pre_REST1} -- \ref{eq_pre_REST5} describe the overhead associated with this scheme.

\vspace{-3ex}

\begin{align}
	t^{j}_{ckpt} &= 0\label{eq_pre_REST1}\\
	t^{j}_{pre} &= \sum_{i=1}^{\#prmpts_{j}}  t^{j}_{pre_i}\label{eq_pre_REST2}\\
        chr^{sys}_{ckpt} &= 0\label{eq_pre_REST3}\\
        chr^{sys}_{pre} &= \sum_{k \in batch\ jobs} t^{k}_{pre} \times nodes_{k}\label{eq_pre_REST4}\\
        chr^{j}_{ckpt} &= 0\label{eq_pre_REST5}
\end{align}

Here, $\#prmpts_{j}$ is the number of times $job\ j$ is preempted, $t^{j}_{pre_i}$ is the time that $job\ j$ (preempted job) has run in its $i$th execution because the job has to start over from the beginning. $nodes_{j}$ is the number of nodes used by $job\ j$. 

{\bf SYS-CKPT:}
This scheme corresponds to the system-level checkpoint support. All BJs are checkpointed \emph{periodically} by the system (without any application assistance), and the checkpoint data (the process memory including the job context) are written to a parallel file system (PFS) for job restart. Running BJs chosen for preemption are killed immediately, and they are resubmitted to the normal queue. When the preempted BJs get to run again, the system resumes them from the latest checkpoint. The system checkpoint interval ($ckpIntv_{sys}$) is universal for all running BJs. Equations~\ref{eq_preckp_sys1} -- \ref{eq_preckp_sys5} describe the overhead incurred by the preempted jobs (in terms of time) and the system (in terms of core-hours).  

\vspace{-3ex}\
  
\begin{align}
	t^{j}_{ckpt} &= \sum_{i=1}^{\lfloor \frac{t^{j}_{runtime}}{ckpIntv_{sys}}\rfloor} \frac{ckpData^{j}_{i}}{BW_{PFS}^{write}} \label{eq_preckp_sys1}   \\
	t^{j}_{pre} &= \sum_{i=1}^{\#prmpts_{j}} \frac{ckpData^{j}_{latest}}{BW_{PFS}^{read}} 
		+ckpTgap^{j}_{i}\label{eq_preckp_sys2}\\               
    chr^{sys}_{ckpt} &= \sum_{k \in batch\ jobs} t^{k}_{ckpt} \times nodes_{k}\label{eq_preckp_sys3} \\
   chr^{sys}_{pre} &= \sum_{k \in batch\ jobs} t^{k}_{pre} \times nodes_{k}\label{eq_preckp_sys4}\\
   chr^{j}_{ckpt} &= 0\label{eq_preckp_sys5}
\end{align}

Here $ckpData^{j}_{i}$ is the amount of data to be checkpointed for job $j$ for $i$th checkpoint, and $t^j_{runtime}$ is the run time of job $j$. $ckpData^{j}_{latest}$ is the amount of data checkpointed in the most recent checkpoint for $job\ j$. $BW_{PFS}^{write}$ and $BW_{PFS}^{read}$ represent the write and read bandwidth of the PFS, respectively. $ckpTgap^{j}_{i}$ is the time elapsed between the time $job\ j$ was checkpointed last and the time job $j$ gets preempted for $i$th preemption.

{\bf APP-CKPT:}
This scheme corresponds to application-level checkpointing. Applications checkpoint themselves by storing their execution contexts and recover by using that data when restarted without explicit assistance from the system. The checkpoint interval ($ckpIntv_{app}^j$) and the amount of data checkpointed ($ckpData^{j}$) change based on the application. Equations~\ref{eq_preckp_app1} to \ref{eq_preckp_app5} describe the overhead incurred by the preempted jobs (in terms of time and core-hours) and the system (in terms of core-hours). 

\vspace{-3ex}

\begin{align}
	t^{j}_{ckpt} &= \sum_{i=1}^{\lfloor \frac{t^{j}_{runtime}}{ckpIntv_{app}^j}\rfloor} \frac{ckpData^{j}_{i}}{BW_{PFS}^{write}} \label{eq_preckp_app1}\\
    t^{j}_{pre} &= \sum_{i=1}^{\#prmpts_{j}} \frac{ckpData^{j}_{latest}}{BW_{PFS}^{read}} 
    + ckpTgap^{j}_{i}\label{eq_preckp_app2}\\
    chr^{sys}_{ckpt} &= 0\label{eq_preckp_app3}\\
    chr^{sys}_{pre} &= \sum_{k\in batch\ jobs} t^{k}_{pre} \times nodes_{k}\label{eq_preckp_app4}\\
    chr^{j}_{ckpt} &=  t^{j}_{ckpt} \times nodes_{j}\label{eq_preckp_app5}
\end{align}

{\bf JIT-CKPT:}
In this scheme, jobs are checkpointed just-in-time (JIT), i.e., right before they get preempted. The premise here is that there is an interaction between the scheduler and the checkpointing module. When the scheduler is about to preempt a job, it informs the appropriate checkpointing module and waits for a checkpoint completion notification before it actually preempts the job. The checkpoint and preemption overhead in this scheme is minimal since there is no need to checkpoint at periodic intervals and there will not be any redundant computation 
(since checkpoint and preemption happen in tandem). As opposed to the other checkpointing schemes described above, RTJs incur an additional delay (the time taken to do JIT checkpoint) in this scheme. Equations~\ref{eq_pre_ckpt1} -- \ref{eq_pre_ckpt5} describe the overhead incurred by the preempted jobs (in terms of time) and the system (in terms of core-hours). 

\vspace{-3ex}
 
\begin{align}
	t^{j}_{ckpt} &= \sum_{i=1}^{\#prmpts_{j}} \frac{ckpData^{j}_{i}}{BW_{PFS}^{write}}\label{eq_pre_ckpt1}   \\
	t^{j}_{pre} &= \sum_{i=1}^{\#prmpts_{j}} \frac{ckpData^{j}_{i}}{BW_{PFS}^{read}}\label{eq_pre_ckpt2} \\
    chr^{sys}_{ckpt} &= \sum_{k\in batch\ jobs} t^{k}_{ckpt} \times nodes_{k}\label{eq_pre_ckpt3} \\
    chr^{sys}_{pre} &= \sum_{k\in batch\ jobs} t^{k}_{pre} \times nodes_{k}\label{eq_pre_ckpt4}\\
    chr^{j}_{ckpt} &= 0\label{eq_pre_ckpt5}
\end{align}

{\bf Summary of the Performance of these Schemes:} We evaluated these schemes using the real-world job logs from Mira. We studied their performance for different percentages of RTJs in the workload. \eunsungnote{The following is related to Reviewer\#1-6 comment}Employing a high-priority queue for RTJ dramatically reduces the slowdown of RTJ (4x or more) when compared with the baseline scheme that treats all jobs equally. However, the absolute values of the average slowdown of RTJs are still around 2.  Preemption is required to bring the average slowdown of RTJs close to 1. Surprisingly, both the high-priority queue and preemptive schemes that favor RTJs benefited BJs also when \%RTJ $\le$ 30.
Further analyses revealed that in addition to RTJ, narrow (jobs that requested a small number of nodes) and short (jobs with short walltime) BJ also benefited significantly from the schemes that favor RTJ. With preemptive schemes, preemption of wide (jobs that requested a large number of nodes) and long (jobs with long walltime) BJ can help narrow and short BJ (in addition to RTJ) through new backfilling opportunities. With high-priority queue, prioritizing RTJ over BJ (and making wide BJ wait) possibly creates additional backfilling opportunities for narrow and short BJ. But the performance of wide and long jobs suffered, which is likely not acceptable for the HPC centers. Thus, we need sophisticated heuristics that protects these (and all other) classes of BJs from significant degradation in performance. We develop improved heuristics in this work to address this issue. First, we present a MILP formulation of the problem to find the optimal solution for small trace, which then can be used to measure how far off the improved heuristics is from the optimal solution. 

\section{MILP-based Real-time Scheduling Algorithm}\label{sec_optimization_algorithm}
In this section, we mathematically formulate the real-time job scheduling problem as a mixed integer linear programming (MILP).
This formulation is used as a baseline for the performance evaluation of our heuristics.

\subsection{Overview of problem formulation} 

The job scheduling problem on modern HPC systems is a multi-dimensional optimization problem. In this study the scheduling problem is a joint optimization problem of improving both \emph{RTJ performance} ($P_{RTJ}$) and \emph{batch job performance} ($P_{BJ}$) with \emph{system constraints} as in Equation~\ref{eq_overall_objective}.
Since RTJs and BJs share system resources (i.e. nodes), the performance improvement of one type of jobs, in general, has negative impacts on the other type of jobs.
In our actual formulation, for the sake of simplicity, we solve the problem with the objective of maximizing BJs' performance while guaranteeing RTJs' performance at least above the predefined threshold. If the formulation fails to solve, the formulation with a looser threshold is tried repeatedly.
\eunsungnote{The following explanation is added according to Reviewer\#4-3 comment.}
For example, if the constraint for RTJs' average slowdown is first set to 1.1 and the formulation fails to solve, that means we cannot find any schedule satisfying 1.1 of RTJs' average slowdown. Then we should try with a looser constraint such as 1.2. The appropriate RTJs' average slowdown can be binary searched, e.g., for the interval of 1.1 and 2.0.
\begin{equation}
\begin{split}
\max & ~P_{RTJ}~and~P_{BJ}\label{eq_overall_objective} \\
\text{ subject to } & job/system~ constraints
\end{split}
\end{equation}

\subsubsection{Real-time job performance ($P_{RTJ}$)}
The performance of RTJs is the first priority. 
Here, we use \emph{job SD} as a performance metric of RTJs. 
Ideally, all the RTJs should have \emph{job SD} close to $1.0$, and a bigger \emph{job SD} means worse performance.
Thus, the term $P_{RTJ}$ in the objective function is the negative sum of all the costs ($Cost_{RTJ}$) of RTJs as in Equation \ref{eq_performance_objective}.

\begin{equation}
P_{RTJ} = -\sum_{j\in Job_{RTJ}} Cost_{RTJ}^{j}		
\label{eq_performance_objective}
\end{equation}
The cost of a RTJ ($Cost_{RTJ}$) can be generalized as a piecewise function of its \emph{job SD} as in Equation~\ref{eq_RTJ_cost_value} if we would like to suppress SDs of jobs around a threshold by sharply increasing the costs beyond the threshold. 
{\small
\begin{equation}
Cost_{RTJ}= 
	\begin{cases}
	f_{s}(SD_{RTJ}) & \text{if } 1.0\le SD_{RTJ}<SD_{thresh}\\
	f_{l}(SD_{RTJ}) & \text{otherwise}  
	\end{cases}
\label{eq_RTJ_cost_value}
\end{equation}
}
In Equation~\ref{eq_RTJ_cost_value}, $f_{s}$ and $f_{l}$ are the cost functions used when \emph{the RTJ slowdown} ($SD_{RTJ}$) is smaller or larger than \emph{the job slowdown threshold} $SD_{thresh}$, respectively. 
\eunsungnote{The following description is modified according to Reviewer\#4-3.}
In this paper, however, we put the objective of minimizing RTJs' cost as a constraint and define the cost of RTJs as the average of RTJs' slowdowns where both $f_s(SD_{RTJ})$ and $f_t(SD_{RTJ}) = SD_{RTJ}$.

\subsubsection{Batch job performance $(P_{BJ})$} 
Based on the checkpoint/preemption scheduling policy, some BJs are checkpointed and will be later restarted if no available resources for RTJs exist due to running BJs. This checkpoint/preemption activity would cause execution overhead to the BJs.

\ignore{
	BJ slowdown \& Incentive. When accommodating RTJs by preempting BJs,  both the running and waiting BJs will suffer slowdown. We define \emph{relative job slowdown} ($SD_{BJ}^{rel}$) as the ratio of the actual \emph{job slowdown} ($SD_{BJ}$) to the \emph{job slowdown} when all the jobs are scheduled as BJs ($SD_{BJ}^{all-batch}$) as in Equation~\ref{eq_bj_rel_slowdown}. 

\begin{equation}
SD_{BJ}^{rel} = \frac{SD_{BJ}}{SD_{BJ}^{all-batch}}
\label{eq_bj_rel_slowdown}
\end{equation}
    
Later the checkpointed BJs will be incentivized with a low charging rate, this reduction in charging price will be compensated by lifting the charge rate for RTJs, and we record this cost as $C_{BJ}^{incentive}$, and illustrated as Equation \ref{eq_cost_incentive}

\begin{equation}
C_{BJ}^{incentive} = \sum_{j\in Job_{BJ}}(Incentive(SD_{BJ}^{j}))
\label{eq_cost_incentive}
\end{equation} 
}

The checkpoint/preemption activity would only generate overhead to BJs. We denote this overhead as $\emph{PreemptCost}_{BJ}$, and it 
is
represented as a function of \emph{checkpoint interval} ($Int_{ckp}$), \emph{checkpoint data size} ($DSize_{ckp}$) and \emph{I/O bandwidth} ($BW_{ckp}$) for checkpoint data storage as in Equation \ref{eq_cost_preempt}. 
We adopt only the just-in time checkpoint scheme for our formulation.
\begin{equation}
PreemptCost_{BJ} = g(Int_{ckp}, DSize_{ckp}, BW_{ckp})
\label{eq_cost_preempt}
\end{equation} 

Like $P_{RTJ}$, $P_{BJ}$ is the negative sum of all the costs ($Cost_{BJ}$) of BJs as in Equation~\ref{eq_performance_objective2}.
$Cost_{BJ}$ is also \emph{the BJ slowdown} ($SD_{BJ}$), which factors in $PreemptCost_{BJ}$.
\begin{equation}
P_{BJ} = -\sum_{j\in Job_{BJ}} Cost_{BJ}^{j}		
\label{eq_performance_objective2}
\end{equation}

\subsubsection{Job/system constraints}
As explained in Section~\ref{subsec_sched_problem}, a job is defined as a tuple of submission time, start time, wall time, run time, job size, and job type (RTJ or BJ). Those attributes of jobs have a relationship among them. For example, start time should be later than submission time.
In addition, when a job is scheduled on computing resources, system resource allocation should be considered.
For example, only consecutive nodes in terms of network connections can be allocated for a job.
Such conditions are imposed as job/system constraints in our formulation.

Besides, regarding big facilities such as Mira at Argonne National Laboratory, the overall system utilization is a critical factor to evaluate the system operating efficiency at the end of every year.
In this study, we introduce average \emph{productive system utilization} ($Util_{productive}$), which is the overall system utilization ($Util_{system}$) subtracted by the utilization caused by checkpoint/preemption overhead ($Util_{overhead}$) as in Equation \ref{eq_util_productive}. 
\begin{equation}
\textit{Util}_{\textit{productive}} = \textit{Util}_{system} - \textit{Util}_{overhead}
\label{eq_util_productive}
\end{equation} 
The checkpoint/preemption schemes can decrease the average productive utilization due to overhead incurred by checkpoint data write/retrieval (and the redundant computation from the time last checkpoint was taken to the time the job was preempted). We consider checkpoint/preemption overhead in both our formulation and heuristics such that unnecessary checkpoint/preemption can be avoided to maximize productive system utilization or restricted under some percentage of overall system utilization.
The detailed constraints and considerations 
are
described in Section~\ref{subsec_formulation_offline_sched} and ~\ref{subsec_heuristic_detail}.

\ignore{
We put the constraint such that the decrease of average \emph{productive utilization} should be less than $X\%$ of the \emph{productive utilization} when all the jobs are considered as BJs ($Util_{all-batch}$).
, as illustrated in Equation \ref{eq_system_constrains}. In this study, we set the relative decrease percentage $X\in(5,10)$. 
\begin{equation}
\textit{Util}_{productive} \geq (1-X\%) \times \textit{Util}_{all-batch}
\label{eq_system_constrains}
\end{equation} 
}

\subsection{Machine-dependent features (IBM BG/Q)}\label{subsec_machine_feature}
Supercomputers may have different architectures 
(for example, node configurations) 
which
affects node allocations when scheduling jobs.

\ignore{
We give a brief overview of the hardware features of IBM BG/Q, and interested readers can refer to \cite{bgq}. Let's take an example of Mira, a IBM BG/Q system, at Argonne National Lab. A midplane is composed of 512 nodes, and two midplanes belong to a rack.
16 racks make up one row, and the total system has 3 rows.}
\rajnote{I think we have the same text as above in section II-c}\eunsungnote{removed}
In this paper, we target Mira, a IBM BG/Q system at Argonne National Laboratory, for our formulation and experiments, but the results can also be applied to other supercomputer systems. 
Mira's various job queues (e.g., prod-capability, prod-short, and prod-long) support different partition sizes, where a partition is a logical boundary
of nodes that can be assigned to a job.
Due to the restrictions of hardware configurations such as midplane, rack, row, and
torus network topology, 
partition sizes supported are 512, 1024, 2048, 4096, 8192, 12288, 16384, 24576, 32768, and 49152.
We define a notion of \emph{partition block} to track the physical location of a partition.
We divide the whole nodes into 512-node blocks which are indexed from partition block 1 to partition block 96(=total number of nodes/512).
The partition block number is used for mixed-integer linear programming in the following sections. \ian{Hard to understand partition block concept, or why it matters.}\eunsungnote{rephrased}

The system memory size of a node is 16 GB. The overall bandwidth of parallel file systems of Mira is 240 GB/s, and we assume 90\% of bandwidth (216 GB/s) can be used for checkpoint/preemption. Mira has one I/O node, which will do I/O service for compute nodes at 4 GB/s, per each 128 compute nodes.

\subsection{Partition-sequence scheduling model} 
Our formulation of the scheduling problem is based on the notions of partition and sequence.
Figure~\ref{fig:seq_part} shows how jobs are scheduled on a machine
which is represented by a graph with 
time on x-axis and partition block number on y-axis, for a partition size of 512 nodes.
The scale of the y-axis is a multiple of 512, and the maximum value of y-axis equals 96 (x 512) in the case of Mira.
The sequence is the order of scheduled jobs on each partition, which means that each partition
counts its sequence number as jobs are scheduled on the partition.
In Figure~\ref{fig:seq_part}, $J_1$ through $J_5$ are scheduled.
$J_1$, $J_2$, and $J_4$ are scheduled on one 512-nodes
whereas $J_3$ and $J_5$ are scheduled on two 512-nodes.
The sequence number of a partition block increases only if a job is assigned on the partition.\rajnote{the above sentence is not clear}\eunsungnote{rephrased}
Thus, the 96th partition block has only one sequence, i.e., $s=1$ for $J_1$.
In the case of the second partition block, three jobs are scheduled on the block, 
which results in sequence numbers ranging from 1 to 3.
Note that the $J_5$ is assigned to different sequence numbers (i.e., 2 and 3) of partition block 1 and partition block 2.

\begin{figure}[h]
	\centering
	\includegraphics[width=1\columnwidth]{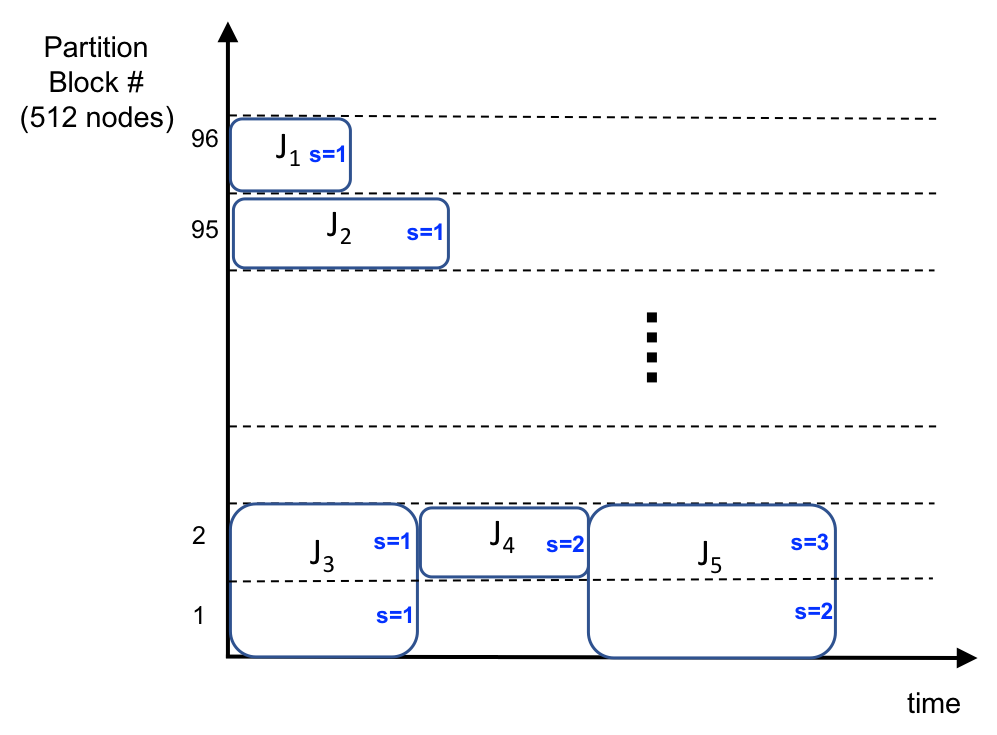}
	\caption{Partition-sequence model for scheduling on parallel machines, especially for Mira.}
	\label{fig:seq_part}
\end{figure}

\ignore{
\begin{figure}[h]
\includegraphics[width=columnwidth]{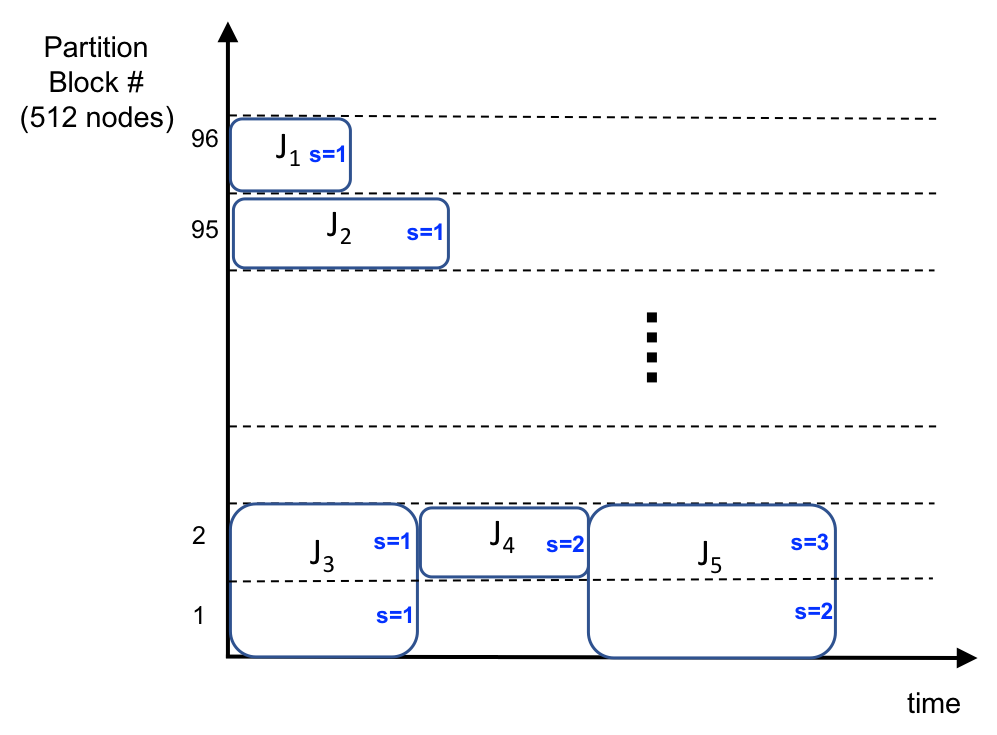}
\caption{Partition-sequence model for scheduling on parallel machines, especially for Mira.}
\label{fig:seq_part}
\end{figure}
}
\subsection{The offline scheduling problem}\label{subsec_formulation_offline_sched}
Our formulation is for offline scheduling of jobs on the Mira supercomputer,
which means all the jobs to be scheduled are known in advance.
There is a mix of BJs and RTJs to be scheduled on the Mira supercomputer.
In addition, we make a few more assumptions in this paper:
\begin{itemize}
\item All the jobs are \emph{rigid}. Thus the number of nodes required for each job is constant and predefined at job submission time. This is consistent with actual job operations in the Mira.
\item Regarding batch-job preemption and restart, we assume that BJs can be preempted by RTJs, and will be restarted from the preemption \rajnote{is it preemption point or last checkpoint?}\eunsungnote{it's preemption point in the formulation} point (with an overhead).
\item RTJs cannot be preempted.
\item Even though heuristics allow a BJ to be preempted multiple times, in our formulation a BJ can be preempted only once.
\end{itemize}

\begin{table*}[bp]
\centering
\begin{minipage}{0.9\textwidth}
\hrule
{\small
\begin{flalign}
&st_{j}, rst_{j} \geq subt_{j}, &j\in J\label{eq_constraint_st1}\\
&st_{j} ex^R_{j,s,p} \geq stSeq_{s,p} + B  (ex^R_{j,s,p}-1),
	&j \in J_R,  s\in S,  p \in P\label{eq_constraint_st2}\\
&st_{j} ex^B_{j,s,p} \geq stSeq_{s,p} + B (ex^B_{j,s,p}-1)
	- B  (exPrmpt_j),
    &j \in J_R,  s\in S, p \in P\label{eq_constraint_st3}\\
&rst_{j} ex^B_{j,s,p} \geq stSeq_{s,p} + B (ex^B_{j,s,p}-1),
	&j \in J_B, s\in S, p \in P\label{eq_constraint_st4}\\
&st_{j} ex^B_{j,s,p} \geq stSeq_{s,p} + B  (\sum_{\substack{s\le si \le t\\: si\in S}} ex^B_{j,si,p}-3)
	+ 2B \ ex^B_{j,s,p} - B \ exPrmpt_j,
	&j \in J_B, s\in S, p \in P\label{eq_constraint_st5}\\
&st_j ex^B_{j,s,p} + rt_j (exRt^B_{j,s,p} \ ex^B_{j,s,p}) + 
	 ovhd_{j}^B (exPrmpt_{j} \ ex^B_{j,s,p}) \leq et_j, 
    &j \in J_B,  s\in S,  p \in P\label{eq_constraint_et1}\\
&rst_j ex^B_{j,s,p} + rt_j(exRt^B_{j,s,p} \ ex^B_{j,s,p}) + 
	ovhd_{j}^B(exPrmpt_{j} \ ex^B_{j,s,p})  &j \in J_B,  s\in S,  p \in P\label{eq_constraint_et2}\\
	&~~~~\leq et_j + B(2-\sum_{\substack{1\le si \le s\\: si\in S}} ex^B_{j,si,p}),&\nonumber
\end{flalign}
}
\end{minipage}
\end{table*}

We first define notations for the rigorous formulation of the offline scheduling problem.
We then formulate the problem as a mixed-integer linear programming (MILP).
\subsubsection{Notation}\label{sssec:notation}
The notations to be used in the job scheduling problem are defined as follows:  
\begin{itemize}
 	\item $\bm{J}$: a total set of jobs,  $j \in J = \{1, ..., N\}$ where $N$ is the number of jobs. 
    \item $\bm{J_{R}}$: a set of RTJs, $j_r \in J_R = \{1, ..., N_R\}$ where $N_R$ is the number of RTJs.
    \item $\bm{J_{B}}$: a set of BJs, $j_b \in J_B = \{1, ..., N_B\}$ where $N_B$ is the number of BJs.
    \item $\bm{B}$: Big number for MILP constraint transformation.
 	\item $\bm{P}$: a set of machine partitions in unit of 512 nodes, $p \in P=\{1, ..., M\}$ where $M$ is the number of partitions in unit of 512 nodes. $M$ equals 96 because the total number of nodes in Mira is 49,152.
    \item $\bm{nodes_j}$: required partition size of job $j \in J$.
    \item $\bm{PB_j}$: a set of possible partition blocks for a job $j$. According to $nodes_j$, $PB_j$ is determined. For example, if $nodes_j=1024$, $PB_j =\{1,...,48\}$.
    \item $\bm{mPartDep[n,blocknr,mapindex]}$: a matrix of partition dependency where $n$ is the number of requested nodes in unit of 512 nodes, $pbnr$ is the element of the possible partition block numbers for $n$, and $mapindex$ is the index in the whole node map. For example, if $n$ is 32(=32*512=16384nodes), $pbnr$ is in the set ${1, 2, 3}$ (The total number of nodes in IB BG/Q is 49,152), and $mapindex$ ranges from 1 to 96 regardless of $n$. When $mPartDep[2,1,1]=1$, this means that the partition block number 1 for the partition size of 1024(=2*512) occupies the first 512-nodes in the system. Likewise, $mPartDep[2,1,2]$ would be 1 since 1024 nodes should occupy two 512-nodes. This variable is needed to detect the overlapping of node allocations of different partition sizes.
    \item $\bm{SDRTJ_{thresh}}$: threshold of slowdown of RTJs.
    \item $\bm{S}$: a set of sequences $s \in S=\{1, ..., T\}$. \eunsungnote{The followng is added for Reviewer\#4-3 comment} For example, in Fig. 1, the partition 1 has two sequences comprising $J_3$ and $J_5$ while the partition 2 has three sequences comprising $J_3$, $J_4$, and $J_5$, which makes $T=3$ in this case.
    \item $\bm{stSeq_{s,p}}$: start time of sequence $s$ of partition $p$.
 	\item $\bm{subt_j}$: submit time of job $j \in J$. 
 	\item $\bm{rt_j}$: runtime of job $j \in J$. not walltime but actual runtime.
 	\item $\bm{st_j}$: start time of job $j \in J$.
    \item $\bm{rst_j}$: restart time of a BJ $j \in J_B$.
    \item $\bm{et_j}$: end time of job $j \in J$.
	\item $\bm{ex_{j,s,p}}$:  binary variable for execution of job $j$ on partition $p \in P$ in the $s$th order.
    \item $\bm{exPb_{j,pb}}$: binary variable for execution of job $j$ on partition block $pb$.
    \item $\bm{exRt_{j,s,p}^B}$: real variable for computation ratio of a BJ across multiple sequences.
    \item $\bm{exPrmpt_{j}^B}$: binary variable for the preemption of a BJ $j \in J_B$.
    \item $\bm{exOvhd_{j,s,p}^B}$: binary variable for checkpoint overhead of a BJ $j \in J_B$ on the sequence $s \in S$ of the partition $p \in P$.
    \item $\bm{ovhd_{j}^B}$: constant value for checkpoint overhead of a BJ $j$. In the formulation, we simply use Equation~\ref{eq_restart_overhead} to compute the overhead, the time taken to checkpoint or restart the checkpointed job, according to Section~\ref{subsec_machine_feature}. Recall that one I/O node per 128 compute nodes can write at 4 GB/s and the parallel file system's maximum I/O throughput is 216 GB/s.
    \begin{equation}
    ovhd_{j}^B = \frac{16*nodes_j}{\min(\frac{nodes_j}{128} \times 4,216)}
    \label{eq_restart_overhead}
    \end{equation} 
\end{itemize}

To help understand binary variables, let us take a look at one of the variables.  $\bm{ex_{j,s,p}^{R}}$ is a binary variable denoting whether the RTJ $j$ on partition $p \in P$ in the sequence $s$ is executed or not, as in Equation~\ref{eq_def_exec1}. Likewise, $\bm{ex_{j,s,p}^{B}}$ is a binary variable for a BJ.\ian{In the following, should RJ be RTJ?}\eunsungnote{fixed}
\begin{equation}
ex_{j,s,p}^{R}= 
\begin{cases}
1 & \text{if } \text{the RTJ } j \text{ is executed on partition } p\\
  & \text{  in sequence } s.\\
0 & \text{otherwise.} 
\end{cases}
\label{eq_def_exec1}
\end{equation}

\subsubsection{Objective}

The objective function in Equation~\ref{eq_performance_objective2} can be translated into minimizing the average SD of BJs as in Equation~\ref{eq_objective} while constraining the average SD of RTJs to a certain extent. The official definition of \textit{job SD} is given in Definition~\ref{def_sd}. 

\begin{defn}\label{def_sd_rt}
\emph{SD} of \emph{RTJ} $j$ $=\frac{(st_j+ rt_j) - subt_j }{rt_j}$. 
\end{defn}

In the case of RTJs, we can expand $et_j$ into $st_j + rt_j$ as in Definition~\ref{def_sd_rt} because RTJs cannot be preempted. 
Constraints are explained in the following sections.
\begin{equation}
\min{ \sum_{j\in J_B}{\frac{et_j-subt_j}{rt_j}} }
\label{eq_objective}
\end{equation}

\subsubsection{Real-time job slowdown constraint}

Equation~\ref{eq_constraint_sd} enforces that the average SD of RTJs is less than or equal to $SDRTJ_{thresh}$.

\begin{equation}\label{eq_constraint_sd}
\frac{1}{N_R}  \sum_{j \in J_R} (st_j + rt_j - subt_j) / rt_j \leq SDRTJ_{thresh}
\end{equation}

\subsubsection{Job start/end time constraints}
\begin{figure}[h]
	\centering	
	\subfigure[]{
    	\includegraphics[height=9ex]{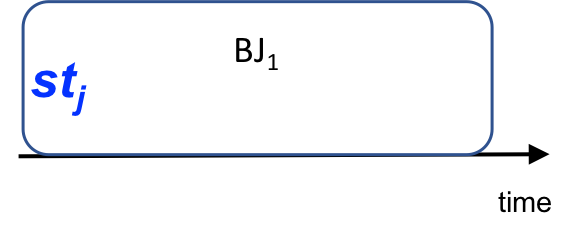}
		\label{fig:st_const_a}
	}
	\subfigure[]{		
    	\includegraphics[height=9ex]{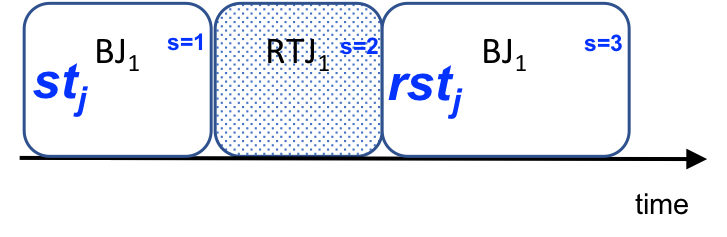}
		\label{fig:st_const_b}
	}
	\caption{Start constraint for a BJ: (a) without and (b) with preemption.}
	\label{fig:st_const}
\end{figure}

Job start/end time constraints ensure that any precedence requirements of jobs are well observed \rajnote{the above sentence is not quite clear}\eunsungnote{rephrased by precedence requirements} and start/end times are correctly associated with sequence start times.
These constraints get complicated due to preemption of BJs.
To model at most once preemption for BJs, a restart time variable for only BJ (i.e., $rst_j$) is introduced.
The start/restart time of a job is greater than or equal to the submission time of the job as in Constraint~\ref{eq_constraint_st1}.
Constraints~\ref{eq_constraint_st2} and \ref{eq_constraint_st3} ensure that a RTJ's start time is greater than or equal to a sequence start time if the job is scheduled on that sequence.
Constraints~\ref{eq_constraint_st4} and \ref{eq_constraint_st5} ensure that a BJ's start/restart time is greater than or equal to a sequence start time if the job is scheduled on that sequence.
Figure~\ref{fig:st_const} shows that Constraints~\ref{eq_constraint_st4} and \ref{eq_constraint_st5} hold for both cases of without preemption and with preemption. 
Here we provide more details on Constraint~\ref{eq_constraint_st5}.
In 
Figure~\ref{fig:st_const_a}, the right-side expression becomes $stSeq_{s,p} (=stSeq_{s,p} + B(1-3) + 2B - 0B)$, in 
the first sequence in Figure~\ref{fig:st_const_b}, the right-side expression 
becomes $stSeq_{s,p} (=stSeq_{s,p} + B(2-3) + 2B - B)$, and finally in 
the third sequence in Figure~\ref{fig:st_const_b}, the right-side expression becomes $stSeq_{s,p}-B (=stSeq_{s,p} + B(1-3) + 2B - B)$, which makes the Constraint~\ref{eq_constraint_st5} always hold true because the start time has nothing to do with the third sequence. Constraints~\ref{eq_constraint_et1} and \ref{eq_constraint_et2} ensure that a BJ's end time is greater than or equal to the BJ's start/restart time.

\subsubsection{Sequence constraints}
Sequence constraints are related to the timing of sequences (i.e., sequence start times) and how jobs are scheduled on sequences.
Constraint~\ref{eq_constraint_seq1} ensures that the start time of the first sequence in any partition is 0.
Constraint~\ref{eq_constraint_seq2} ensures that the start time of a sequence is less than or equal to the start time of the next sequence.
Constraint~\ref{eq_constraint_seq3} and \ref{eq_constraint_seq4} ensure that if a RTJ is scheduled on the sequence $s$, the end time of the job should be less than the start time of sequence $s+1$, because RTJs cannot be preempted.
Constraints~\ref{eq_constraint_seq5} -- \ref{eq_constraint_seq7} ensure that if a BJ is scheduled on the sequence $s$, the end time of the BJ composed of the run time and the preemption overhead should be less than the start time of the next sequence $s+1$. \rajnote{the above sentence is not clear.}\eunsungnote{rephrased}
Constraints~\ref{eq_constraint_seq6} and \ref{eq_constraint_seq7} are more complicated: 
as with Constraints~\ref{eq_constraint_st4} and \ref{eq_constraint_st5}, they must allow for both the BJs that get preempted and the BJs that do not get preempted, \rajnote{I rephrased the above sentence. Eunsung, please check.}\eunsungnote{checked}
as shown in Figure~\ref{fig:st_const}.

{\footnotesize
\begin{alignat}{2}
&stSeq_{1,p} = 0, &p \in P\label{eq_constraint_seq1}\\
&stSeq_{s,p} \leq stSeq_{s+1,p}, &s\in S\backslash T, p \in P\label{eq_constraint_seq2}\\
&(stSeq_{s,p} + rt_j) ex^R_{j,s,p} &j \in J_R,  s \in S\backslash T, p \in P\label{eq_constraint_seq3}\\
	&\qquad\leq stSeq_{s+1,p}, &\nonumber\\
&(st_j + rt_j) ex^R_{j,s,p} \leq stSeq_{s+1,p},
	&j \in J_R,  s \in S\backslash T,  p \in P\label{eq_constraint_seq4}
\end{alignat}
}

\begin{table*}[bp]
\centering
\begin{minipage}{0.9\textwidth}
\hrule
{\small
\begin{align}
&(stSeq_{s,p} + rt_j exRt^B_{j,s,p})  ex^B_{j,s,p} + ovhd_{j}^B (exPrmpt_{j} ex^B_{j,s,p} )  \leq stSeq_{s+1,p},  &j \in J_B, s \in S\backslash T, p \in P\label{eq_constraint_seq5} \\
&(st_j + rt_j exRt^B_{j,s,p}) ex^B_{j,s,p} + (exPrmpt_j  ovhd^B_j) ex^B_{j,s,p} \leq &j \in J_B, s \in S\backslash T, p \in P\label{eq_constraint_seq6}\\
	&\qquad stSeq_{s+1,p} + B(2-\sum_{\substack{s\le si \le t\\: si\in S}} ex^B_{j,si,p}) + B(exPrmpt_j -1), &\nonumber\\
&(rst_j + rt_j exRt^B_{j,s,p}) ex^B_{j,s,p} + ovhd_{j}^B (exPrmpt_{j} ex^B_{j,s,p}) \leq &j \in J_B, s \in S\backslash T, p \in P \label{eq_constraint_seq7}\\
	&\qquad stSeq_{s+1,p} + B(2-\sum_{\substack{1\le si \le s\\: si\in S}} ex^B_{j,si,p}), \nonumber 
\end{align}
}
\end{minipage}
\end{table*}

{\footnotesize
\begin{alignat}{2}
    &\sum_{\substack{s \in S, pb \in PB_j,\\ p \in P}} \big(mPartDep[\textstyle{\lceil \frac{nodes_j}{512} \rceil},pb,p]& j\in J_R\label{eq_constraint_assign1}\\
    	&\qquad \times (ex^R_{j,s,p} exPb_{j,pb})\big) = \lceil \textstyle{\frac{nodes_j}{512}} \rceil, &\nonumber\\
    &\sum_{s \in S, p \in P}  ex^R_{j,s,p}  = \lceil \textstyle{\frac{nodes_j}{512}} \rceil, &j\in J_R\label{eq_constraint_assign2}\\
    &\sum_{\substack{s \in S, pb \in PB_j,\\ p \in P}} \big(mPartDep[\textstyle{\frac{nodes_j}{512}},pb,p]& j\in J_B\label{eq_constraint_assign3}\\
    	&\qquad\times (exRt^B_{j,s,p} exPb_{j,pb})\big) = \lceil\textstyle{\frac{nodes_j}{512}} \rceil, &\nonumber\\
    &\sum_{s \in S, p \in P} exRt^B_{j,s,p} = \lceil \textstyle{\frac{nodes_j}{512}} \rceil, &j\in J_B\label{eq_constraint_assign4}\\
    &\sum_{j \in J_R} (ex^R_{j,s,p}) + \sum_{j \in J_B} (ex^B_{j,s,p}) \leq 1, &s \in S, p \in P& \label{eq_constraint_assign5}\\
    &\sum_{s \in S} (ex^R_{j,s,p}) \leq 1, &j \in J_R, p \in P \label{eq_constraint_assign6}
\end{alignat}
}

\subsubsection{Job assignment constraints}
A job is assigned to a partition block at some point of time.
A BJ can be preempted whereas a RTJ can not be preempted.
Thus, a preempted BJ is assigned to the same partition block at two different sequences.
Our formulation allows for only one preemption, so as to avoid more complicated decision variables and additional constraints.

Constraint~\ref{eq_constraint_assign1} ensures that the number of nodes assigned to a job $j$ is same as the requested number of nodes.
Constraint~\ref{eq_constraint_assign2} ensures that a RTJ occupies the requested number of nodes.
Constraints~\ref{eq_constraint_assign3} and \ref{eq_constraint_assign4} ensures that a BJ occupies the requested number of nodes.
Constraint~\ref{eq_constraint_assign5} ensures that at most one job can run at a certain sequence on any partition.
Constraint~\ref{eq_constraint_assign6} ensures that a RTJ can run on a certain partition at most one time throughout all sequences.

\subsubsection{Preemption constraints for batch jobs}
We allow for at most one preemption for a BJ, and also take
into account the overhead of preemption/restart.
Constraint~\ref{eq_constraint_preempt1} ensures that the partial execution ratio of a BJ is greater than or equal to the threshold, which is 0.1 in this equation.
Constraint~\ref{eq_constraint_preempt2} enforces that the number of preemptions across partitions are the same.
Constraint~\ref{eq_constraint_preempt3} ensures that the number of allocated partitions matches the request number of nodes, for both with and without preemption.
Constraint~\ref{eq_constraint_preempt4} matches batch ratios with scheduled sequences on partitions.

{\footnotesize
\begin{alignat}{2}
   &\frac{ex^B_{j,s,p}}{10} \leq exRt^B_{j,s,p} \leq ex^B_{j,s,p}, &j \in J_B, s \in S,\label{eq_constraint_preempt1}\\
   && p \in P\nonumber\\
   &\sum_{s \in S} ex^B_{j,s,p} = exPrmpt^B_j+1,&j \in J_B,  p \in P\label{eq_constraint_preempt2}\\
   &\qquad\text{if } \sum_{s \in S} ex^B_{j,s,p} > 0,  &\nonumber\\
   &\sum_{\substack{s \in S,\\ p \in P}} ex^B_{j,s,p} = \textstyle{\lceil \frac{nodes_j}{512} \rceil}(exPrmpt^B_j+1),  &j\in J_B\label{eq_constraint_preempt3}\\
   &\sum_{s \in S} (exRt^B_{j,s,p} \times ex^B_{j,s,p}) = \text{1 or 0}, &j \in J_B, p \in P\label{eq_constraint_preempt4}
\end{alignat}
}

\ignore{
\begin{equation}\label{eq_constraint_preempt2}
\sum_{s \in S} (exRt^B_{j,s,p} \times ex^B_{j,s,p}) =
\begin{cases}
1 & \text{if } \sum_{s\in S} ex^B_{j,s,p} > 0,\\
0 & \text{Otherwise} \\ 
\end{cases}
\forall j \in J_B, \forall p \in P
\end{equation}
}

\ignore{
\begin{equation}\label{eq_constraint_preempt5}
   \sum_{s \in S} ex^B_{j,s,p} \leq MaxPreempt+1;
\end{equation}
}

\subsubsection{Additional constraints for partitions (specific to IBM BG/Q)}
Only one partition block among possible partition blocks depending on $nodes_j$ is assigned to a job.
This constraint can be formulated as follows.
\begin{equation}\label{eq_constraint_part1}
\sum_{pb \in PB_j} exPb_{j,pb} = 1, \forall j \in J
\end{equation}

%

%
%
%

\subsubsection{Transformation of formulation}
Some constraints in the formulation such as Constraint~\ref{eq_constraint_preempt3} need a transformation to be regular equations or inequalities. We use Big M method~\cite{bradley_applied_1977} for this purpose, and one constraint can be split into one or more equations and inequalities. For example, a \emph{if-then-else} statement can be split into two or more equations and inequalities.
Also for quadratic terms such as a multiplication of binary variable(b) and a real variable(r) can be solved by introducing a new variable z($=b\times r$) with inequalities.
We do not show the complete transformations due to the space limitation.

\subsection{Complexity Analysis}
The time complexities of linear programming problems depend on the number of variables and constraints whereas the time complexities of mixed integer linear programming problems also depend on the number of integer variables. 
As described in Section~\ref{sssec:notation}, the maximum number of a certain type of variable (i.e. $ex_{j,s,p}$) among all variables is $N\times M \times T$ where $N$ is the number of jobs, $M$ is the number of partitions, and $T$ is the number of sequences. The thing making it worse to compute the optimal solution in a reasonable time is those variables that are binary. 
Table~\ref{tbl:cplex_time} shows the number of binary variables after preprocessing the problem \rajnote{what is pre-solve?}\eunsungnote{rephrased by preprocessing} and run times of solving formulations when using IBM CPLEX~\cite{cplex} as the problem size is growing. We ran these sample experiments on a 32-core machine. We can see the run time grows exponentially with the size of the problem, and it takes up to 12 hours when the number of jobs is seven, and the partition size and the sequence size are 16 and 5, respectively.
Thus, we evaluate our heuristics with our formulation when the number of jobs is seven. The details are described in Section~\ref{sec_evaluation}.

\ignore{
\begin{table}
	\centering
	\includegraphics[width=0.9\columnwidth]{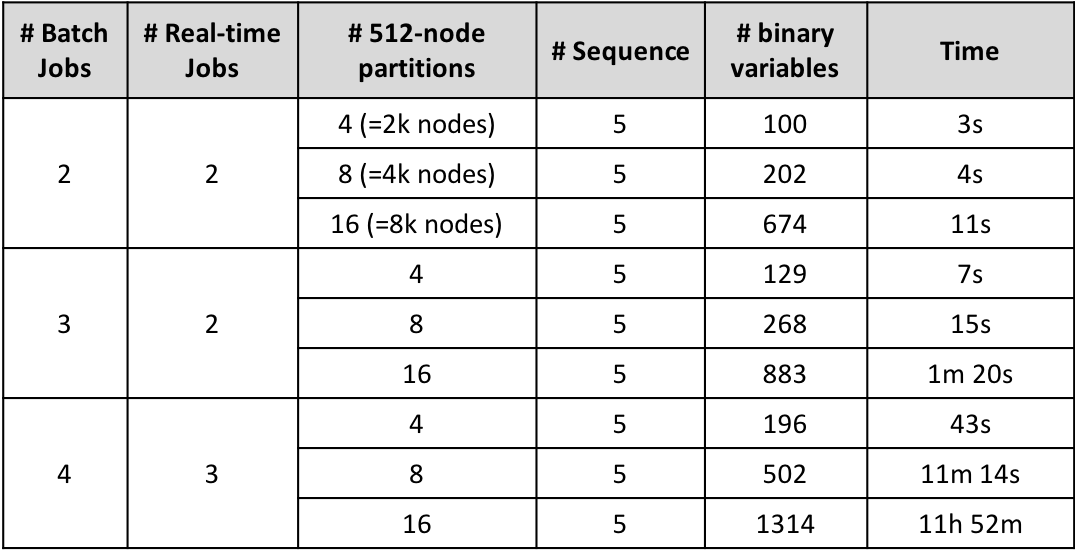}
	\caption{Run times of solving formulation using IBM CPLEX~\cite{cplex}.}
\end{table}
}

\begin{table}
	\centering
    \caption{Run times of solving formulation using IBM CPLEX~\cite{cplex}.}
	\label{tbl:cplex_time}
    {\footnotesize
    \begin{tabular}{ |c|c|c|c|c|c| } 
     \hline
     \multirow{2}{2.5em}{\# BJs} & \multirow{2}{3em}{\# RTJs} & \# 512-node & \multirow{2}{2em}{\# seq.} & \# binary & \multirow{2}{3em}{Time} \\ 
      &  & partitions &  &  variables &  \\
     \hline\hline
     \multirow{3}{2.5em}{2} & \multirow{3}{2.5em}{2} & 4(=2k nodes) & 5 & 100 & 3s\\
     && 8(=4k nodes) & 5 & 202 & 4s\\
     && 16(=8k nodes) & 5 & 674 & 11s\\
     \hline
     \multirow{3}{2.5em}{3} & \multirow{3}{2.5em}{2} & 4 & 5 & 129 & 7s\\
     && 8 & 5 & 268 & 15s\\
     && 16 & 5 & 883 & 1m 20s\\
     \hline
     \multirow{3}{2.5em}{4} & \multirow{3}{2.5em}{3} & 4 & 5 & 196 & 43s\\
     && 8 & 5 & 502 & 11m 14s\\
     && 16 & 5 & 1314 & 11h 52m\\
     \hline
    \end{tabular}
    }
	
\end{table}

\section{Novel Scheduling Heuristic}\label{sec_heuristic}

As summarized in Section~\ref{sec_basic_scheduling}, our previous work~\cite{wang_supporting_2017} showed that simple as well as some sophisticated heuristics to support RTJs can be detrimental to certain classes of batch jobs. Here we build on that work to develop a new heuristic that addresses the issues with the heuristics described in Section~\ref{sec_basic_scheduling}.
To develop our scheduling heuristic, we made the additional changes to the modified Cobalt scheduling heuristic in \cite{wang_supporting_2017} as follows.
\begin{itemize}
\item
Split jobs into a low priority and a high priority queue to give preferential treatment to jobs that our scheduler deems as high priority. \rajnote{what is this for? for prioritizing certain batch jobs over other batch jobs?} \eunsungnote{According to Algorithm 1, the qualified RTJs into a high priority queue and others into a low priority queue.}
\samnote{Eunsung is correct, all batch jobs are in low priority queue. The high priority queue is for realtime jobs that have reached a slowdown threshold and so we want to run them ASAP}
\item
Created a weighted cost scheduling algorithm for high priority queue jobs.
\end{itemize}

\subsection{Estimated Slowdown}
Before going into more details of the scheduling heuristic, it is essential to understand the primary metric used for the heuristic's decision-making process: estimated slowdown. A job submitted to an HPC system has 
a walltime, which is a user submitted time for a job that corresponds to the maximum possible runtime for the job. 
The actual runtime, which is the time that the job actually runs in the HPC system and can only be computed after the job completes.

When scheduling jobs, our heuristic computes a job's estimated slowdown and then uses this value to make decisions. For jobs that haven't started running yet, we define the estimated slowdown (ESD) of a job as below.

{\footnotesize
\begin{equation} 
ESD = \frac{job.wallTime + currentTime - job.submitTime}{job.wallTime} \label{eq:estimatedSlowdown1}
\end{equation}
}
For jobs that are currently running, we define ESD as
{\footnotesize
\begin{equation} 
ESD = \frac{job.estimatedEndTime - job.submitTime}{job.wallTime} \label{eq:estimatedSlowdown2}
\end{equation}
}
where estimatedEndTime of a job is computed as follows.
{\footnotesize
\begin{align} 
&job.estimatedEndTime = \\ \nonumber
&job.wallTime + currentTime - job.runningTime \label{eq:estimatedEndTime}
\end{align} 
}
Thus, a job's ESD is a way of estimating what a job's slowdown would be if the job were to (start now, if not started already, and) run for completion. 

Note that the equation for the ESD is different from the BSD (bounded slowdown) equation that is used to evaluate the performance of the scheduling algorithms. The equation for ESD  uses a job's walltime instead of runtime since in the real world we don't know a job's actual runtime until after it completes, and thus we use the job's walltime as a substitute.

\begin{algorithm}
{\small
\DontPrintSemicolon 
$threshold \gets highPriorityQueueThresholds[job.category]$\;
\For{$job$ \textbf{in} $availableJobs$} {    
    \If{$job.realtime$ = True \textbf{and} $job.estimatedSlowdown > threshold$} {
        $highPriorityQueue$.add($job$);
    }
    \Else{
        $lowPriorityQueue$.add($job$);
    }
}
\If{ len($highPriorityQueue$) $ > 0 $} {
    \For{$job$ \textbf{in} $highPriorityQueue$.sortBySlowdown()} {
        weightedCostSchedulingAlgorithm($job$); // Algorithm \ref{algo:weightedCostAlgorithm}}
}
\ElseIf{ len($lowPriorityQueue$) $ > 0 $} {
    \For{$job$ \textbf{in} $lowPriorityQueue$} {
        defaultSchedulingAlgorithm($job$);
    }
}
\caption{{\sc Scheduling Heuristic}}
\label{algo:schedulingHeuristic}
}
\end{algorithm}

\subsection{Heuristic Details}\label{subsec_heuristic_detail}
We use two queues:
one low priority and one high priority. The low priority queue contains BJs and also RTJs below a configurable RTJ slowdown threshold;
these jobs cannot preempt any currently running jobs. 
The high priority queue contains all other RTJs;
these jobs can preempt running BJs under certain conditions. 
\samtextold{ For our experiments we set the slowdown threshold to be 1.10.} 
\samtextnew{ For our experiments, we set a different RTJ slowdown threshold for each job category. To set these threshold values, we computed the median slowdown values for the log files under a baseline simulation (all jobs are treated as batch jobs) and then set the threshold value to be 50\% of this value. We then set limits on these values with the minimum value for threshold being 1.1 and the maximum value being 2.0. We picked these values since we felt they represented a reasonable range of slowdown values for RTJs. For the Mira log used in the simulations the RTJ slowdown threshold values are: 1.1 for narrow-short jobs, 1.1 for narrow-long jobs, 2.0 for wide-short jobs and 1.6 for wide-long jobs. The details of the different job categories are discussed in more detail in Section \ref{ref_experiment_setup_job_categories}. Although under this approach the RTJ slowdown threshold values are hard coded into the simulator, it would be relatively easy to implement a system that updated these values dynamically. Such as computing a running average of slowdown values over the previous week.} 
\samnote{Should I also include the threshold values for the CEA Curie logs as well?}
\samnote{Is this enough justification for how we picked the threshold values or should I add more?}

We note that the high priority queue in this scheme is different from the high-priority queue \rajnote{please check the abbreviation}\eunsungnote{checked. just used the full name} scheme in \cite{wang_supporting_2017}. The high priority queue in this scheme only has RTJs that have reached a slowdown threshold, while he high-priority queue scheme in \cite{wang_supporting_2017} contains all RTJs.  \samnote{What we had before was incorrect so I revised it. We will change this for the final version of the paper and update it accordingly: "In the high-priority queue scheme, low priority queue jobs are only considered if the high priority queue is empty. Here the jobs in the low priority queue are attempted to be scheduled after attempting to schedule all the jobs in high priority queue even if all the jobs in high priority queue are not scheduled." } \rajnote{please check if the above statement is true}\eunsungnote{I agree. not matched with Algorithm 1}
We use a weighted cost job scheduling (WCJS) algorithm for scheduling jobs in high priority queue, which is discussed in the next section. \rajnote{what about the jobs in low priority queue?}\eunsungnote{scheduled based of default Cobalt scheduling policy; score computed based on \#nodes, waittime, etc.}
\samnote{the details for the low priority queue jobs are in the next section. Should I move them here?}
The complete algorithm is shown in Algorithm~\ref{algo:schedulingHeuristic}.


\begin{algorithm}
{\small
\DontPrintSemicolon 
\KwIn{A job, $job$, in the high priority queue}
$partitions \gets$ getPartitionsBySize($job.nodes$)\;
\For{$p$ \textbf{in} $partitions$} {
    \If{$p.available =$ True} {
        runJob($job$, $p$)\;
        \Return{}\;
    }
}
$possiblePartitions \gets []$\;
\For{$p$ \textbf{in} $partitions$} {
    \If{$p.preemptRealtime() =$ True} {
        \textbf{continue};
    }
    \ElseIf{$p.preemptBiggerJob() =$ True} {
        \textbf{continue};
    }
    \ElseIf{$p.preemptThreshold() =$ True} {
        \textbf{continue};
    }
    $possiblePartitions$.add($p$)
}
$bestPartition \gets None$\;
$bestScore \gets \infty$\;

\For{$p$ \textbf{in} $possiblePartitions$} {
    $totalScore \gets 0$\;
    
    \For{$childJob$ \textbf{in} $p.getChildJobs()$} {
        $score \gets childJob.nodes$\;
        $score \gets score \times childJob.slowdownFactor()$\;
        $score \gets score \times childJob.checkpointFactor()$\;
        $score \gets score \times childJob.timeRemainingFactor()$\;
        $totalScore \gets totalScore + score$\;
    }
    \If{$totalScore < bestScore$} {
        $bestPartition \gets p$\;
        $bestScore \gets totalScore$\;
    }
}
runJob($job$, $bestPartition$)\;
}
\caption{{\sc Weighted Cost Job Scheduling}}
\label{algo:weightedCostAlgorithm}
\end{algorithm}

\subsection{Weighted Cost Job Scheduling Algorithm}

At each iteration of the scheduler, our scheduling heuristic separates all available jobs (i.e., jobs that are ready to be run but are not currently running) into the two queues. BJs are automatically placed in the low priority queue. For RTJs, an estimated slowdown value is computed and jobs with values below the threshold are placed in the low priority queue, while the remainder are placed in the high priority queue.

Next, the scheduler attempts to schedule the queued jobs. Our algorithm only schedules low priority queue jobs if the high priority queue is empty. Otherwise, the algorithm schedules the high priority queue jobs. \rajnote{isn't this a problem? I suppose this will starve the BJs} \samnote{As we discussed, this could be a problem, I will look into this after we submit the paper} 
For low priority queue jobs, our scheduling heuristic scores the jobs based on walltime, queuetime, and number of nodes and then uses these scores to sort the jobs in the queue. Then for each job in the newly sorted queue, the heuristic looks for an available partition for the job to run on, and if a matching partition is found, it runs the job.  
\samnote{Rephrased description of scheduling for low priority queue above - please review. Removed: "uses the default Cobalt scheduling algorithm which does not preempt currently running jobs, and so we do not discuss the details of the algorithm here. Please refer to the Cobalt scheduler documentation for more information."} \rajnote{I think we should avoid referring to cobalt and just provide a brief description here.}

When scheduling high priority queue jobs, our heuristic uses a custom weighted cost job scheduling algorithm that has the ability to preempt currently running BJs that meet certain criteria. Thus, only RTJs with an estimated slowdown value above a certain threshold can preempt jobs in the system, and BJs can never preempt other jobs. 

When a job is placed in the high priority queue, the weighted cost scheduling algorithm tries to identify the best partition in the system for high priority queue jobs to run in. For this, the algorithm takes all partitions that are of the right size for the job, i.e., the smallest partitions that are still bigger or equivalent in size to the job. For example, a 1000 node job would want a 1024 node partition. \rajnote{partitions with at least 1024 nodes or partitions with only 1024 nodes?} \eunsungnote{the latter}

As shown in Algorithm \ref{algo:weightedCostAlgorithm} lines 1-5, if there is an appropriately sized partition that is available (no job is running on the partition), then there is no need to preempt a running job, and the job is run on this partition. Otherwise, it identifies all eligible partitions.
Eligible partitions are the ones that
\begin{enumerate}
    \item 
    Do not have any RTJs running.
    \item
    Do not have any jobs bigger than the preempting job.
    \item
    Do not have BJs that have a slowdown that exceeds a certain threshold based on the job's size and walltime.
\end{enumerate}
\rajnote{I phrased the criterias above. please check}
\samnote{looks good}

Accordingly, the justification for these criteria are as follows -- preempting RTJs will incur additional delays to them and thus will make them as non real-time; preempting Bigger (in terms of number of nodes) BJs will have a hard time getting back in after being preempted; Preempting BJs that already have a high slowdown value could result in unreasonably high slowdowns for them.

\samtextnew{The BJ slowdown thresholds used in criteria 3 above are different from the RTJ slowdown thresholds used to determine when a RTJ enters the high priority queue. However, they are computed in a similar fashion. We use the median slowdown value for each job category from baseline simulations (all jobs treated as BJs), and then set the BJ slowdown threshold to be 150\% of this value. For the Mira log used in the simulations, the BJ slowdown threshold values are: 1.5 for narrow-short jobs, 1.5 for narrow-long jobs, 17.3 for wide-short jobs and 4.8 for wide-long jobs. }
\samnote{Should I also include the threshold values for the CEA Curie logs?}
\samnote{Is this enough justification for how we picked the threshold values or should I add more?}

It is important to note that partitions are organized in a parent-child hierarchical structure based on the topology of the supercomputer. A parent partition is a superset of any of its child partitions. For example, 1024-nodes partition is composed of two 512-nodes children partitions. You can only run a job on the parent partition if you preempt all of the child partitions' jobs. Thus, for all of the correctly sized partition it is necessary to check their child partitions for criteria 1 and 3, and their parent partitions for criteria 2. 
As long as one partition in the system meets the criteria, then the job will be run right away. If not, the job waits in the queue until a valid partition becomes available.

Next, the algorithm (lines 17-27) computes a weighted cost value for all partitions that meet the criteria. The cost is calculated by summing together scores for all of the jobs in the partition that would be preempted by the new job. The score for each preempted job consists of the number of nodes for the preempted job multiplied by three different factors, each of which has a different weight. The three factors are the slowdown of the preempted jobs, the last checkpoint time of the preempted jobs (if the experiment is using an interval checkpointing scheme), and the percentage of time remaining for the preempted jobs. 

Accordingly, the rationale for these factors are preempting jobs with high slowdown values is bad for performance, preempting jobs that checkpointed recently is better than preempting jobs that are not checkpointed recently, and we want to avoid preempting jobs that are almost done. Although these factors and their weights appear somewhat arbitrary, they are the result of extensive testing done to determine which factors and weights produced the best results. Once this weighted cost is computed for all available partitions, the partition with the lowest cost is selected, all jobs running in the partition are preempted, and the new job is run.

\section{Simulation-based Evaluation}\label{sec_evaluation}
In this section, we present the detailed methodology for a simulation-based evaluation and the evaluation results. \eunsungnote{The following sentence is added to address Reviewer\#4-3 comment.} We also compare the results of heuristics with those of MILP-based formulation to evaluate the performance and possible improvement of heuristics.

\subsection{Experimental Setup} 
\subsubsection{Job Categories} \label{ref_experiment_setup_job_categories}
To understand simulation results better, we divided the jobs into four categories: two classes for the number of nodes (narrow and wide) and two classes for the runtime (short and long). The criteria used for classification is as follows:
\begin{itemize}
\item Narrow: number of nodes used is in the range [512, 4096] inclusive (note that the number of nodes allocated on Mira is a multiple of 512).
\item Wide: number of nodes used is in the range [4608, 49152] inclusive.
\item Short: jobs with runtime $\le$ 120 minutes. 
\item Long: jobs with runtime $>$ 120 minutes.
\end{itemize}
Consequently, there are four job categories -- narrow-short, narrow-long, wide-short, and wide-long.
\subsubsection{Workload Trace} 
\samtextold{ To get realistic simulation results, we used the real job trace log collected from the Mira system in 2018.  
In this paper, we used two week-long logs. 
To establish the baseline for future comparison, we first capture the performance of 
batch scheduling algorithm employed at Mira by running both RTJs and BJs as BJs on the Qsim simulator.}
\samtextnew{ To get realistic simulation results, we used the real job trace logs collected from two different systems: the Mira system in 2018, and the CEA Curie from 2012. 
The CEA Curie log is from the Parallel Workloads archive and more info about the log and CEA Curie system can be found on the Parallel Workloads archive's website. 
Since the CEA Curie log was created by a different scheduler and is based on a different sized system, some minor preprocessing was needed to be done to make it optimized for the simulations. To do this, we normalized the job size for each job in the log as shown in the equation below.
Since the Mira log was created by the same scheduler that we are using for these simulations, the log did not need to have any preprocessing done.
For both logs, we simulated 2 weeks of jobs, but then only computed our simulation results based on the middle 7 days of the simulations to ensure that our simulations were as close to real world scenarios as possible. }
\begin{equation}
jobSize_{normalized}  = \frac{jobSize_{original} \times Mira.node\_count}{CEA\_Curie.node\_count}
\label{eq_cost_preempt}
\end{equation} 

\samnote{Should I add a reference to the Parallel workloads archive here?}

\subsubsection{Job Variation} \label{ref_experiment_setup_job_variation}
\samtextold{
To fully evaluate the performance of all the scheduling schemes under different amount of RTJs, we randomly choose $R\%$ (RTJ percentage) of jobs in the experimental trace log, and set them as RTJs, with the rest $(100-R)\%$ as BJs. Here in our experiments we use $R \in\{5,10,15,20\}$. Experimental results were average from 10 random sample groups for each $R$ value. }

\samtextnew{
To fully evaluate the performance of all the scheduling schemes under different amounts and different types of RTJs, we ran experiments with different R\% (RTJ percentage) of jobs and with different methods for selecting RTJs. Here in our experiments we use $R \in\{5,10,15,20\}$. Experimental results were average from 10 random sample groups for each $R$ value. 
We used 3 different methods for generating RTJs for the experiments. 
}
\begin{enumerate}
    \item 
    Default: we randomly choose $R\%$ (RTJ percentage) of jobs in the experimental trace log, and set them as RTJs, with the rest $(100-R)\%$ as BJs.
    \item
    90 minute walltime threshold: same as the default method except that we only allow jobs in the trace log with a runtime of less than 90 minutes to be RTJs. This method is designed to reflect situations in which RTJs have similar to characteristics to BJs except that they are by nature shorting running jobs with a higher slowdown sensitivity.
    \item
    APS Log Jobs: all jobs in the original trace log are BJs. We generate RTJs for the experiment by using log data taken from the APS (Advanced Photon Source), which a particle accelerator at Argonne National Lab. The log used contains the jobs run to process data for one of the research stations at APS. Running APS jobs on the Mira supercomputer one was of the primary motivators for this research and so it made sense to use this data to simulate RTJs on the Mira system. However, the APS log data is incomplete, and it only contains job runtimes or job sizes. To use the APS log data as RTJs in our experiments, we randomly select a number of jobs from the log (based on the number of jobs in the batch log being used, and the \% of RTJs for the experiment) 
\end{enumerate}

\samtextnew{ The numbered list above is new too, but I couldn't color it }


\subsubsection{Checkpointing Methods}

To evaluate our scheduling heuristic discussed above, we implemented three different checkpointing methods: application checkpointing with 5\% overhead (APP-CKPT-5\%), application checkpointing with 10\% overhead (APP-CKPT-10\%), and just in time checkpointing (JIT-CKPT). 

The application checkpointing method is based on the idea that a job's total checkpointing overhead should be less than some percentage of the job's entire runtime. For implementing this, the scheduler computes the time to checkpoint each job and then uses this value to calculate the number of times each job can be checkpointed so that the total checkpointing overhead is less than or equal to a certain percentage (either 5\% or 10\%) of the walltime (as we don't know the actual runtime of a job before the job completes).
The scheduler, then evenly distributes that number of checkpoints throughout the walltime of the job. For example, if the checkpoint overhead is 10\%, a 120-minute job that takes 5 minutes to checkpoint can be checkpointed two times without exceeding the checkpoint overhead limit (12 minutes). In this case, the job will checkpoint every 40 minutes of runtime. 

\ignore{
The just-in-time checkpointing method consists of only checkpointing a job if it is about to be preempted. When the scheduler identifies a BJ to be preempted by a RTJ, the BJ is stopped, it checkpoints while the RTJ waits, and after checkpointing completes, the RTJ is then run. The rationale behind this method is that the majority of checkpoints are unnecessary and so to avoid this wasted time, only checkpoint when you need to, i.e., a BJ is going to be preempted. Although this can hurt the performance of RTJs since they have to wait while the BJ checkpoints, on average, checkpointing a job doesn’t take that long (average checkpointing time for our experiments is under 10 minutes).}

\subsubsection{Performance metrics for evaluation} 
\label{ref_subsec_metrics_1}
We analyze the performance of various scheduling algorithms in terms of slowdown and turnaround time of the jobs.
\rajnote{I am not sure what exactly we mean by both individual job and overall system performance. So I removed it.}\eunsungnote{that may be because we used system utilization metric in JSSPP.}
To establish the baseline for future comparison, we first capture the performance of 
batch scheduling algorithm employed at Mira by running both RTJs and BJs as BJs on the Qsim simulator.

\subsection{Scheduling Heuristic Simulation Results}
\begin{figure*}
	\centering
	\includegraphics[width=0.75\textwidth]{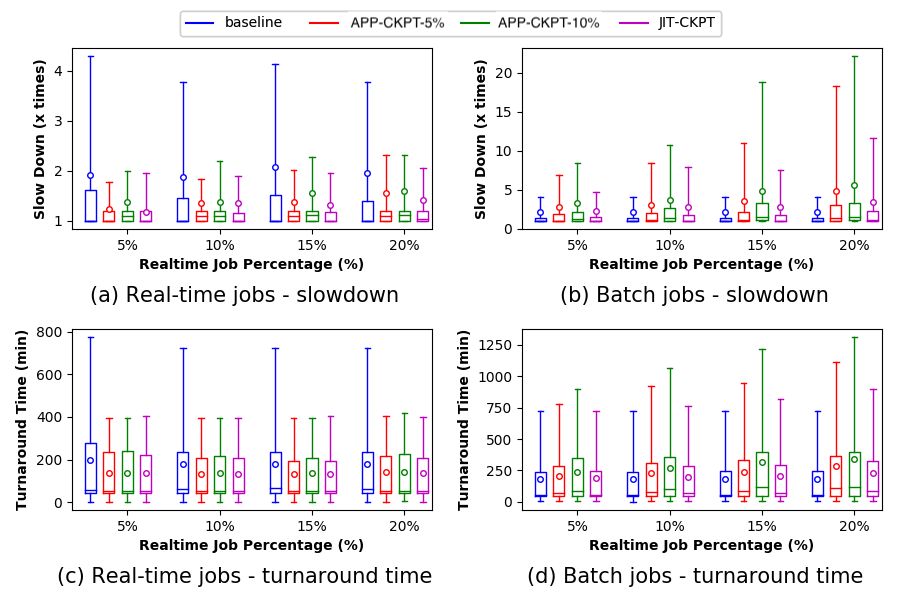}
	\caption{Average slowdown and turnaround time --- overall}
	\label{fig:result-all}
\end{figure*}

\begin{figure}[htb]
	\centering
	\includegraphics[width=1\columnwidth]{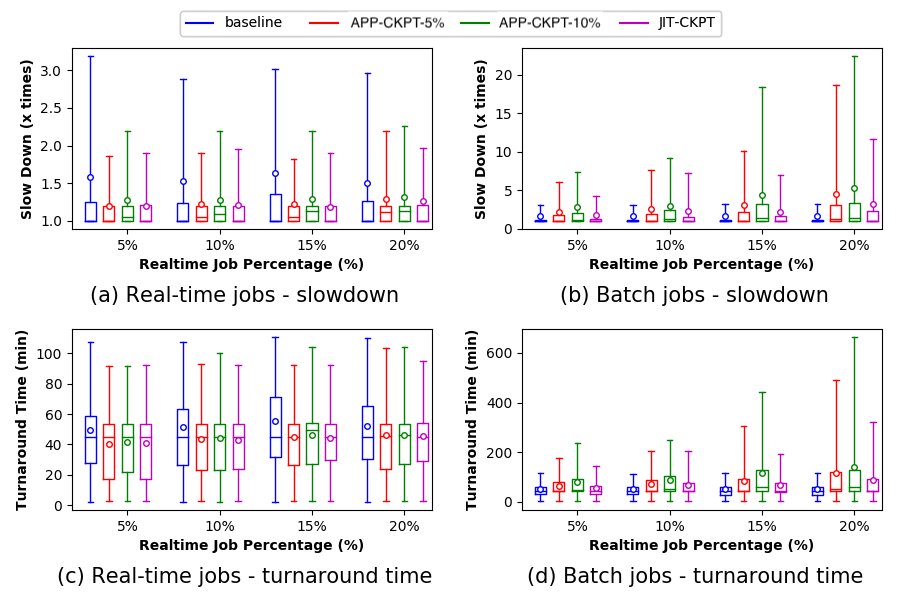}
	\caption{Average slowdown and turnaround time --- narrow-short}
	\label{fig:result-narrow-short}
\end{figure}

\begin{figure}[htb]
	\centering
	\includegraphics[width=1\columnwidth]{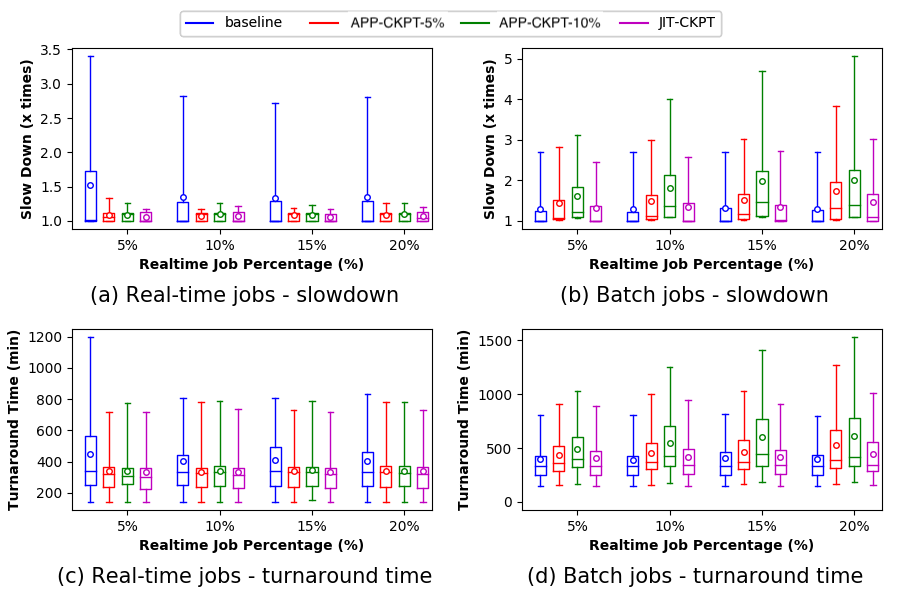}
	\caption{Average slowdown and turnaround time --- narrow-long}
	\label{fig:result-narrow-long}
\end{figure}

\begin{figure}[htb]
	\centering
	\includegraphics[width=1\columnwidth]{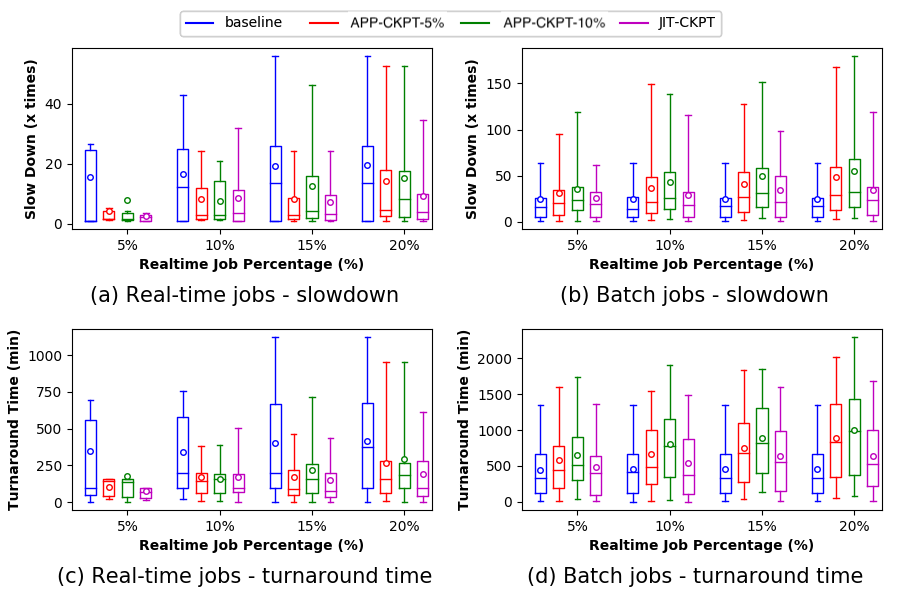}
	\caption{Average slowdown and turnaround time --- wide-short}
	\label{fig:result-wide-short}
\end{figure}

\begin{figure}[htb]
	\centering
	\includegraphics[width=1\columnwidth]{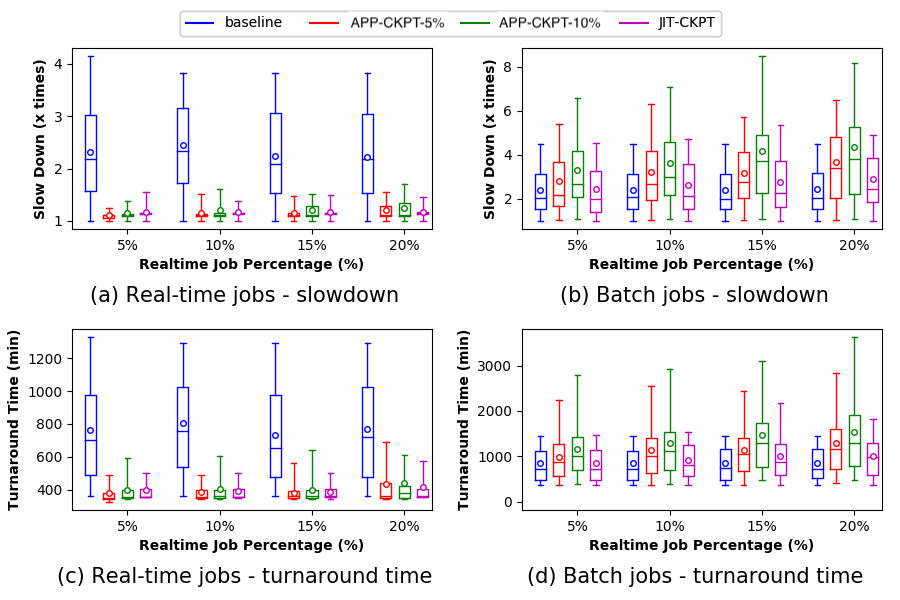}
	\caption{Average slowdown and turnaround time --- wide-long}
	\label{fig:result-wide-long}
\end{figure}

We illustrate our results with a series of figures, which contain the simulation performance results for one week-long trace. Due to space constraints, we do not include the results for another week-long trace; we note that the trends for the other log were similar to that of the trends for the trace presented here. There are five figures for results. Fig.~\ref{fig:result-all} contains results for all jobs, and Fig.~\ref{fig:result-narrow-short} through \ref{fig:result-wide-long} contain results for each of the four job categories, narrow-short, narrow-long, wide-short, wide-long respectively. 

In all figures, the top and bottom row contains a boxplot for the bounded slowdown values and the turnaround times respectively. Each row is divided into two columns, with the left column showing the results for real-time jobs and the right column showing the results for BJs. Each subplot contains 16 different boxplots in groups of 4. The four groups of boxplots correspond to the four different \% of RTJs simulated. Within each group is four boxplots which correspond to the four different checkpoint heuristics evaluated. The four different checkpoint heuristics are color-coded as indicated in the legend at the top of each figure, and the coloring is consistent across all subplots and figures. As a result, the x-axes, indicating the \% of RTJs, are the same for all plots, but y-axes, indicating either the slowdown or turnaround time, are different for each subplot. 

The bottom and top whiskers of the boxplots correspond to the 5\% and 95\% values respectively. The bottom, middle and top lines of the boxes for the boxplots correspond to the 25\%, 50\% (median) and 75\% values. The dots of the boxplots represent the average value. For both metrics provided, slowdown and turnaround time, lower values are better. The lowest possible slowdown value for all jobs would be 1 since this would mean that a job does not have to wait at all. The lowest possible turnaround time value for a job is the job's runtime and thus is job dependent. We decided to present the results in the form of boxplots since we felt it necessary to show not only the average results from the simulation but also the range of values, especially, the upper range or maximum values. We chose to put the whiskers at 5\% and 95\% in an attempt to prevent outliers from skewing the results. However, there are many cases in the figures in which it is difficult to see all lines of a boxplot, in particular, the bottom whisker, 25\% line and 50\% line. In these cases, we felt that these values were all negligibly small enough relative to the other values in the plot that this loss of granularity was okay since it enables showing the upper range of values. 

We first discuss the high-level trends that are evident in all jobs as shown in Fig.~\ref{fig:result-all}, before we move on to discussing the trends and unusual behaviors in the job category figures. To start with, RTJ slowdown values and turnaround times are across all figures lower than BJ values. 
Our custom scheduling heuristics perform better compared to the baseline for RTJs, but they perform worse for the BJs. 
This is to be expected, since the performance improvements experience by the RTJs with our scheduling heuristics come to some extent at the expense of the BJs' performance. However, since there are so many more BJs than RTJs, the ratio of performance variation is much higher for the RTJs than it is for the BJs. I.e., the RTJs' performance improvement is much larger than BJs' performance decline. 

When comparing our three scheduling heuristics, we found that APP-CKPT-5\% performed better than APP-CKPT-10\% for both the cases of the median value and the 95\% value. This implies that the time saved with the extra checkpoints in APP-CKPT-10\% does not make up for all of extra overhead involved in doing those extra checkpoints compared to the APP-CKPT-5\%, which has only half the number of checkpoints\rajnote{half of the heuristics?} \samnote{sorry meant checkpoints instead of heuristics}. 
Similarly, we found that the JIT-CKPT  performed better than the APP-CKPT-5\%  in terms of both the median value and the 95\% value. Like the previous comparison, this implies that it is better to checkpoint only when needed as in the JIT-CKPT heuristic instead of periodically checkpointing as in the APP-CKPT-5\% heuristic. 
This is likely because the overhead cost of periodically checkpointing in the APP-CKPT-5\% heuristic outweighs any performance benefits of periodically checkpointing. \rajnote{This does not make sense. What do we mean by time saved with the checkpoints here in comparison to JIT?}. \samnote{rephrased. Let me know if still not clear. Could potentially remove sentence.} There are a few cases in the figures where the 95\% value performance for the APP-CKPT-5\%  is actually better than that for the JIT-CKPT. However, these cases are only for the RTJs, and are likely caused by the the RTJs having to wait until the JIT checkpointing of BJs is completed before starting to run. \rajnote{not convincing. I believe the reason for this is the extra wait time RTJs experience in JIT. RTJs have to wait until the JIT checkpointing of BJ is done before starting.} \samnote{good point, I revised}

The trends are unclear when comparing the performance for the different percentages of RTJs. In Fig.~\ref{fig:result-all}, when \% RTJs increases, the performance decreases for BJs, but there is not any clear performance difference for RTJs. The performance decline experienced by the BJs corresponds with the fact that having more RTJs in the system means that more jobs have higher priority and can preempt the BJs in the system. This explanation also explains the near constant performance seen by the RTJs since even as \% RTJs increases, the extra RTJs have the same high priority as the rest of the RTJs. 
However, unlike the different checkpoint heuristics, the goal of testing with different \% RTJs is less about determining which \% RTJs has the best performance and more about 
how high of \% RTJs that the scheduler can handle will still delivery adequate performance for both job types. \rajnote{do we have any clarity on how high of \% RTJs that the scheduler can handle will still delivery adequate performance for both job types} \samnote{I was unsure how to handle this. We never did any experiments with values higher than 20\% though for the final version of the paper I could. I also wasn't sure if we thought that 20\%  RTJs could still delivery adequate performance for both job types. So I'm not sure what to put here}

Next, we evaluate the results of different job categories.
There are significant performance variations in terms of slowdown and turnaround time between the four different job categories for both BJs and RTJs. These trends are evident not only in our scheduling heuristics but also in the baseline heuristic, implying that the performance variations are the result of the underlying nature of the jobs and not artifacts of our scheduling heuristics. 

It is important to note that when comparing job performance between the different job categories, the slowdown is a more natural metric than turnaround time since slowdown is normalized to a job's runtime, while turnaround time is not. Thus, the performance variations seen in turnaround time across job categories is mainly due to the different job lengths. 

Fig.~\ref{fig:result-narrow-short} through \ref{fig:result-wide-long} show that narrow-long jobs have the best slowdown performance in terms of both the median and 95\% values.  Narrow-long RTJs using the JIT-CKPT heuristic have near real-time median slowdown values (1.08 when there are 10\% RTJs), and narrow-long BJs using the same heuristic have similar performance to the baseline heuristic ($\sim$1.40). It makes sense that narrow jobs would perform better than wide jobs since it is much easier for the scheduler to accommodate and find space for the narrow jobs, thus allowing them to run sooner and have lower slowdowns.

Narrow-short RTJs have a similar median performance to the narrow-long jobs since they benefit from their smaller job size as well. However, they have significantly worse 95\% performance, which is likely caused by the higher variation between walltime and runtime for the narrow-short jobs. There are a number of narrow-short RTJs with runtimes that are proportionally much shorter than their walltimes. Since the slowdown threshold for RTJs is based on walltime, a job may end up with a higher slowdown value than we expect it to. For example, a job with a walltime of 60 minutes that only runs for 10 minutes, might wait for 12 minutes in the scheduler since we think this will give it a slowdown of 1.2; however, due to its short runtime its actual slowdown will be 2.2 even though it only waited for 12 minutes. Thus, even though these jobs have relatively short wait times in the scheduler queue, their short runtimes cause them to have much higher slowdowns compared to the rest of the narrow-short RTJs. Narrow-long RTJs are less affected by this slowdown skewing since they have longer walltimes and therefore lower proportional variation between walltime and runtime. Another factor for the large value for 95\% performance is the checkpointing time involved in JIT-CKPT when a batch job needed to be checkpointed and preempted before a RTJ can be run. Especially in the case of extra short RTJs, this checkpointing time can cause a significant increase in waittime and thus resulting slowdown value. \rajnote{Not convincing. why would a handful of jobs impact 95th percentile performance. Why is bounded slowdown metric not helping addressing the issue of extremely short jobs skewing the results?} \samnote{I rewrote this, let me know if you still disagree. The bounded slowdown is based on walltime, so for jobs with a much shorter runtime than walltime, they will have much higher slowdowns}  

Narrow-short BJs have worse median performance compared to narrow-long BJs which we also expect is due to the shorter nature of the narrow-short jobs which causes them to be more affected by wait-times even if those wait-times are well within an acceptable range for standard HPC uses. 

Wide-long RTJs also have a similar performance to the narrow RTJs for our scheduling heuristics. 
However, the median performance gap between the baseline and the heuristics for wide-long RTJs is higher compared to narrow RTJs, which means our scheduling heuristic offers wide-long RTJs a relatively bigger performance increase compared to the narrow jobs.
\samnote{agree with Raj's comment below. This seems out of place and unneccesary. Should I just remove the next 2 sentences?}\eunsungnote{I removed}
\ignore{
Under the baseline heuristic, the narrow jobs already had excellent slowdown performance, and so the scope for \% performance improvement for these jobs using our scheduling heuristics is much smaller than that for wide jobs. For example, with 10\% RT job and JIT-CKPT, the average performance improvements for narrow-short and wide-short RTJs were 26\% and 38\%, respectively.} \rajnote{we have been discussing the performance of narrow-long and wide-long but we suddenly switch to narrow-short and long-short} 

Wide-short jobs are the outliers when it comes to performance. This is best demonstrated by the significantly high slowdown values in Fig.~\ref{fig:result-wide-short}(a) and (b) compared to the other job categories. However, these higher slowdowns are represented in the baseline heuristic as well as our scheduling heuristic, which implies that they are due to the differing characteristics of the jobs themselves and not our scheduling heuristic. In fact, as mentioned above, the wide-short RTJs experience vastly improved performance with our scheduling heuristics compared to the baseline. 
One obvious question that this raises is why are the wide-short jobs slowdown values so much higher compared to the other job categories. We hypothesis that similar to the performance difference experienced between the narrow-short and narrow-long jobs mentioned above, the wide-short jobs are much more sensitive to wait times than the wide-long jobs due to their shorter runtimes. This issue is further compounded by the fact that since they are wide jobs, it is much harder for the scheduler to find space for them, which causes them to wait significantly longer in the queue compared to narrow-short jobs. 

Though again, because performance variation is experienced in the baseline as well as our scheduling heuristic, our primary goal is not to get all job categories to have the same performance. Instead, our goal is to have our scheduling heuristics give RTJs the best performance possible while still ensuring BJs have comparable performance to their baseline benchmark for each job category. 

One abnormality in the wide-short category is that the JIT-CKPT heuristic has a higher 95\% slowdown value for RTJs than the APP-CKPT-5\% heuristic for the 10\% RTJs case. Since there are so few jobs of this category in the simulation, this is the result of a small number of wide-long RTJs (~3) with significantly higher slowdowns values that skew the 95\% value. These higher slowdowns are likely the result of the simulation randomly selecting some wide-long jobs with a large difference between their' walltimes and runtimes to be RTJs, which as discussed for the narrow-short jobs can cause higher slowdown values than expected by the scheduler. In addition, since these jobs have such short runtimes, any time required to checkpoint and then preempt running batch jobs in order to run these RTJs can have a large impact on the RTJs' slowdown values, even though the checkpointing times are not actually that long. 
\rajnote{here again I think the reason is the extra wait time for RTJs with JIT because the checkpoint of BJs happen after the RTJs are picked for execution.} \samnote{I rewrote what I had before, but I'm pretty sure though that this is the result of 3 jobs having high slowdown values while the rest of the values are much lower. I believe these significantly higher slowdown values are the result of a large variation between the jobs runtimes and walltimes. I didn't have a chance to prove this, but I think this is a more likely reason to account for such large slowdowns, since I don't think JIT checkpointing takes so long that it causes that large of slowdowns. Let me know if you still disagree and I can rewrite it to reflect what you put in your comment.}

The average simulation values follow many of the same trends as the median. 
The average values are higher than the median values (as shown in the plots by the average dot being above the median line in nearly all of the box plots). 
This is likely the result of there being a number of large outlier values that make the average value higher than the median value; this also corresponds with the high top whiskers seen for many of the boxplots.
Regarding average values, the JIT-CKPT still performs the best among the three custom scheduling heuristics in nearly all of the different simulation configurations. JIT-CKPT performs better than the baseline heuristic for RTJs and performs worse for BJs; however, as is the case for the median values, the performance degradation for the BJs is much smaller than the performance improvement seen by the RTJs. 

Take for example the simulations using JIT-CKPT and 10\% RTJs, which we consider as a very reasonable scenario for practical realization (It is relatively easier to convince the HPC administrators to support a small percentage of RTJs)
The results show that across all the categories, the JIT-CKPT significantly reduces RTJs slowdown by 35\% (from 1.94 in the baseline to 1.25), while only increasing BJs slowdown by 10\% (from 2.26 to 2.51). Furthermore, when broken into the four job categories, 3 of the categories have a BJs slowdown increase under 10\%. Narrow-short jobs are the only category that doesn't, and in this case, the BJs slowdown values go from 1.68 to 1.94 which we argue is still a reasonable slowdown value for these types of jobs. All four categories experience significant improvement in slowdown performance for RTJs ranging from 20\% for narrow-short jobs, since the baseline value was already very low, to 60\% for wide-long jobs (from 3.0 to 1.2), to 80\% for wide-narrow jobs (from 24.40 to 5.06). Thus, the categories that have the highest slowdown values for RTJs are also the categories that experience the largest performance improvements.  One unusual data point to point out is that in the wide-short figure (Fig.~\ref{fig:result-wide-short}) \rajnote{can we be more specific about the figure} \samnote{added figure number}, the average RTJ slowdown value for APP-CKPT-10\% with 5\% RTJs is higher than even the top whisker. This is likely simply due to a very small number of data points (<5), and thus it only takes one large value to bring the average much higher than the boxplots. 

For the sake of space, we don't discuss in details the average turnaround time performance details; however, as shown in the figures the trends and relative performance variations are similar to those experienced by the average slowdown values. 

\eunsungnote{For Rreviewer\#4-8}
There may exist batch jobs which are unable to checkpoint. In such cases, the changes in the heuristic would be minimal since  the heuristic can handle such un-checkpointable workloads as non-preemptible batch jobs, which will worsen the slowdowns of real-time jobs depending on the percentage of un-checkpointable workloads.

\subsection{Comparison with MILP-based formulation}
To evaluate the performance of our heuristic, the Weighted Cost Job Scheduling Algorithm, against the near-optimal scheduling, we compare with our optimal offline scheduling formulation for small problem sizes.
Recall that the MILP-based formulation has limitations in scheduling large problem sizes.
The major differences of the optimal formulation from the WCJS algorithm are:
\begin{itemize}
\item The formulation obtains the optimal solution.
\item The formulation schedules jobs given in advance (i.e., offline scheduling) while the WCJS schedules jobs in an online fashion.
The offline scheduling definitely shows better performance than the online scheduling.
\item The formulation permits at most one preemption of a BJ whereas the WCJS has no limitation on that.
\item The formulation allows a BJ to preempt other BJs whereas the WCJS allow only RTJs to preempt BJs.  
\item The formulation strictly restrict the upper limit of SD of RTJ to $SDRTJ_{thresh}$(=1.2).
\end{itemize}

We  randomly  generated  20  job  logs  with  5  BJs  and  2 RTJs  where  the  size  of  the  available  nodes  is  8096  nodes, and  accordingly,  the requested  node  size  of  jobs  are  also randomly  chosen  in  the  set  of \{512,  1024,  2048,  4096,  8192\}. Table~\ref{table:compare_form_WCJS}  shows  the  comparison of the formulation and WCJS heuristic results.  The  WCJS  heuristic with JIT-CKPT shows 69\% and 38\% worse than the optimal formulation in terms of BJ and RTJ slowdowns, respectively. This suggests that there are still opportunities for improvement for  the  WCJS  such  as  taking  multiple  jobs  in  wait  queues into account when scheduling.

\begin{table}  \renewcommand{\arraystretch}{1.1}
	\centering
	\caption{Performance comparison of the optimal formulation and the WCJS heuristic.}\label{table:compare_form_WCJS}
	\centering
	\begin{tabular}{ |c|c|c| }
		\hline 
		Algorithm & \emph{BJ SD} & \emph{RTJ SD}\\
		\hline \hline
		Formulation & 1.50 & 1.08\\
        WCJS & 2.54 & 1.50 \\
		\hline
	\end{tabular}
\end{table}

\section{Conclusion}\label{sec_conclusion}
We evaluate how existing HPC platforms can accommodate real-time jobs in this paper. We first present the results of applying existing basic techniques to this problem and show mathematical formulations as mixed-integer linear programming to compute near-optimal solutions. We then propose the weighted-cost job scheduling heuristic for practical deployment to overcome the high time complexity of mixed-integer linear programming. The results show that with 10\% real-time job percentages, just-in-time checkpointing method combined with our heuristic could improve the slowdowns of real-time jobs by 35\% while limiting the increase of the slowdowns of batch jobs to 10\%. Based on the comparison results with mixed-integer linear programming, we believe there is still a room for improvement. Our results are promising for existing HPC platforms to accommodate real-time workflow applications, the newly emerging applications if the platforms support checkpointing methods and the scheduler algorithms deploy our proposed heuristics.

\section*{Acknowledgment}
We thank the Argonne Leadership Computing Facility at Argonne National Laboratory for providing the Mira trace log used in this study. 

\bibliographystyle{abbrv}
\bibliography{ref}

\end{document}